# Exciton Bohr radius of lead halide perovskites for photovoltaic and light-emitting applications

**Hyun Myung Jang,[1][*][†] Kyung Yeon Jang,[2][*] Song Hee Lee,[1] Jinwoo Park,[2] Tae-Woo Lee[1,2,3,4][†]**

[1] Research Institute of Advanced Materials (RIAM), Seoul National University, Seoul 08826, Republic of Korea. [2] Department of Materials Science and Engineering, Seoul National University, Seoul 08826, Republic of Korea. [3] Institute of Engineering Research, Soft Foundry, Interdisciplinary Program in Bioengineering, Seoul National University, 1 Gwanak-ro, Gwanak-gu, Seoul 08826, Republic of Korea. [4] SN Display Co. Ltd., Seoul 08826, Republic of Korea.

[*] These authors contributed equally to this work: Hyun Myung Jang and Kyung Yeon Jang

[†] Corresponding author. Email: jang0227@snu.ac.kr (H.M.J.); twlees@snu.ac.kr (T.W.L.)

**ABSTRACT**

**Exciton Bohr radius ($a_B$) and exciton binding energy ($E_b$) of metal halide perovskites are two prime quantities in their applications to both light-emitting diode displays and photovoltaic devices. We develop a reliable theoretical method of simultaneously finding $a_B$ and $\varepsilon_r^c$ (dielectric constant) based on the net exciton energy above the bulk band gap. It is estimated that $a_B$ under the dielectric confinement is substantially smaller than $a_B$ in the absence of dielectric confinement: 4.36 *nm vs.* 5.61 *nm* in the case of CH$_3$NH$_3$PbBr$_3$. We attribute the enhanced $a_B$ to variations of $\varepsilon_r^c$ and the electron-hole correlation energy. We also develop a simple method of finding $E_b$ based on the same net exciton energy. Using this, we attribute the well-known difference in $E_b$ between organic bromide perovskites and iodide counterparts to $\varepsilon_r^c$ and explain that iodide perovskites are more suited than bromide counterparts in photovoltaic applications, which require smaller $E_b$ for efficient charge-carriers transport.**

# INTRODUCTION

Metal halide perovskites (MHPs) demonstrate unusually high-power conversion efficiencies up to 20% and beyond as outstanding photovoltaic devices (*1,2*). In addition, they uniquely exhibit superior electro-optical properties as light emitters for light-emitting diodes (LEDs) (*3-5*). Their narrow emission linewidths, quantified by full width at half maximum (FWHM), enable high color purity in LED displays with FWHM ≤ 20 nm and color gamut covering ~100% of the Rec.2100 standard. These attributes surpass the characteristics of organic emitters and inorganic quantum dot (QD) emitters (*6*). The use of colloidal perovskite nanocrystals (PNCs) is an effective strategy for increasing the photoluminescence quantum efficiency (PLQE) of MHP crystals (*7-10*). Organic ligands adsorbed on the nanocrystal surface (*11,12*) tend to limit the crystal growth of the PNC cores and thus enhance their dispersion stability in solution. Since MHPs have low exciton binding energies, restricting the PNC size with ligands helps prevent thermal ionization of electron-hole pair (or exciton) (*5,7*).

According to our experimental investigation on the PNC size-dependent photoluminescence (PL) properties of $MAPbBr_3$ colloidal dispersion, the PL wavelength (PL-$\lambda$) is weakly dependent on the PNC size down to the modal radius $R$ of ~5 nm (diameter $D$ of ~10 nm) but decreases noticeably below this critical value (Fig. 1), where MA denotes methylammonium ($CH_3NH_3$) organic group. This weakly size-dependent PL-$\lambda$ for $R \geq$ ~5 nm is accompanied by the size-insensitive PL-emission linewidth with the FWHM at ~23 nm (Fig. 1). The linewidth increases rapidly below this critical size (*12*). In addition to these two spectral features, it was further suggested (*13*) that the size-dependent PLQE of $MAPbBr_3$ colloidal dispersion shows its broad maximum at the PNC size slightly bigger than this critical size ($R >$ 5 nm; green vertical zone in Fig. 1). The rapid decrease in the PL-$\lambda$ (*i.e.*, enhanced PL blue shift) for $R \leq$ 5 nm indicates that a strong quantum-confinement, which accompanied by the band-gap widening, becomes pronounced at $R =$ ~5 nm. It was proposed that this critical radius corresponds to the exciton Bohr radius ($a_B$) since a strong quantum-confinement effect, as reflected in the degree of PL blue shift, becomes pronounced at this critical size (*13*).

In the case of $MAPbBr_3$ colloidal dispersion, the PNC radius slightly larger than 5 nm is promising since the size-independent PL-$\lambda$ characteristics are expected above this critical size, while maintaining both high PLQE and color purity owing to a small constant FWHM value (Fig. 1). Thus, $a_B$ is an important parameter to deduce optimum conditions for exhibiting promising PL properties that include (i) size-independent PL wavelength, (ii) PL linewidth, and (iii) high PLQE. However, the reported values of $a_B$ for $MAPbBr_3$ are widely scattered**:** 1.36 nm (*14*), 1.45 nm (*15*), 2.2 ~ 4.0 nm (*16*), 4.38 nm (*17*), 4.7 nm (*18*), 5.0 nm (*13,19*). $a_B$ is defined as the effective distance between electron (*e*) and hole (*h*) with the reduced mass of $\mu_o$ in each *e-h* pair. Thus, $a_B$ in a medium of the relative dielectric permittivity $\varepsilon_r$ (dielectric constant hereafter) is defined as (*20*)

$$a_B = \varepsilon_r \left(\frac{m_o}{\mu_o}\right) a_o = \frac{4\pi\varepsilon_o\varepsilon_r\hbar^2}{\mu_o e^2} \qquad (1)$$

where $m_o$ is the free-electron mass ($9.109 \times 10^{-31} \, kg$), $\mu_o$ is the effective reduced mass of exciton, and $a_o$ denotes the Bohr radius for the *1s* orbital of a hydrogen atom ($= 4\pi\varepsilon_o\hbar^2/m_o e^2 = 0.0529 \, nm$). In the case of nanocrystal (NC) at $R_c$, the onset size of the PL blue shift, $\varepsilon_r$ in Eq. 1 should be replaced by $\varepsilon_r^c$ which is the dielectric constant of the NC core at $R_c$. It is known that a reliable value of $\mu_o$ can be obtained by careful data fitting of the magneto-optical absorption spectroscopy (*21*): in the case of MAPbBr$_3$, $\mu_o = 0.117 m_o$. Other studies also reported similar values: $\mu_o = 0.12 m_o$(*13*) and $\mu_o = 0.13 m_o$(*22*). Thus, the scattered values of $a_B$ can be attributed mainly to the difficulty in locating the appropriate value of $\varepsilon_r^c$ under the context of *e-h* interaction in a given exciton.

Considering this difficulty, we developed a reliable theoretical method of simultaneously finding optimum values of $a_B$ and $\varepsilon_r^c$ starting from the net exciton energy above the bulk band gap which, in turn, is based on the Brus-Takagahara-type standard form of Hamiltonian operator (*23,24*). According to our self-consistent iterative calculations, $a_B$ under the dielectric confinement, which is suitable for a colloidal dispersion, is substantially smaller than $a_B$ in the absence of dielectric confinement which is relevant to a single crystalline material. Additionally, we developed a simple method of finding the exciton binding energy ($E_b$) starting from the same net exciton energy. This method of evaluating $E_b$ considers both the classical Coulombic attraction and the quantum *e-h* correlation effect. Thus, the present method is a substantial improvement over the current method (*25,26*) that considers only the Coulombic interaction between electron and hole in each exciton pair. By applying the present theory to the well-known difference in $E_b$ between organic bromide perovskites (*e.g.*, MAPbBr$_3$) and organic iodide perovskites (*e.g.*, MAPbI$_3$), we attribute this difference mainly to the enhanced $\varepsilon_r^c$ value in organic iodide perovskites (*21*). This explains that organic iodide perovskites (*e.g.*, MAPbI$_3$ and FAPbI$_3$) are more appropriate than bromide counterparts in photovoltaic applications, which require smaller $E_b$ for efficient charge-carriers transport.

## RESULTS

### Energy of a confined exciton under dielectric confinement

The energy of a dielectrically confined exciton above the bulk band gap ($\Delta E_g$) is the starting point of our analysis and discussion of $a_B$ and $E_b$. Therefore, we will briefly discuss $\Delta E_g$ under the dielectric confinement. Detailed systematic derivation and explanations of $\Delta E_g$ with and without the dielectric confinement were presented in our earlier publication (*19*). If the dielectric constant of the surrounding medium is lower than that of the dispersed NC cores, the surrounding medium tends to exert a repulsive force on the NC core, resulting in a dielectric-confinement effect on the core. As described in the **INTRODUCTION**, this situation corresponds to the measurement of a PL spectrum from the colloidal NCs in a surrounding low-permittivity medium (*19,27-30*). Under this condition, the effective Coulomb interaction coefficient ($C_1'$) for the (*s-s*)-type ground state of exciton is substantially reduced from the well-known unmodified coefficient $C_1$ (= 1.786) which corresponds to the purely quantum confinement in the absence of any dielectric-confinement effect (*20,31,32*). In the below, we will show that the value of $C_1'$ can be obtained numerically as a function of the permittivity ratio, $\varepsilon_r^c/\varepsilon_r^s$, where $\varepsilon_r^c$ and $\varepsilon_r^s$ denote the dielectric constants (permittivities) of the NC core and the surrounding medium, respectively. To this end, we first consider the Hamiltonian operator ($\mathcal{H}_{NC}$) of the confined exciton within the spherical NC core with radius $R$ (Fig. 2A). For quantitative analysis, we use the following Brus-Takagahara-type standard form of $\mathcal{H}$ in the presence of dielectric-confinement effect (*23,24*):

$$\mathcal{H}_{NC} = -\frac{\hbar^2}{2m_e}\nabla_e^2 - \frac{\hbar^2}{2m_h}\nabla_h^2 - \frac{e^2}{4\pi\varepsilon_o\varepsilon_r^c |r_e - r_h|} + \frac{e^2}{4\pi\varepsilon_o(2R)}\sum_{n=0}^{\infty}(\bar{\varepsilon}_n)^{-1}\left\{\left(\frac{r_e}{R}\right)^{2n} + \left(\frac{r_h}{R}\right)^{2n}\right\} -$$

$$\frac{e^2}{4\pi\varepsilon_o(R)}\sum_{n=0}^{\infty}(\bar{\varepsilon}_n)^{-1}\left(\frac{r_e r_h}{R^2}\right)^n P_n(\cos\theta_{eh}) + V(r_e, r_h) \tag{2}$$

where $r_e$ and $r_h$ denote the coordinates of an electron and a hole, respectively, $m_e$ and $m_h$ are their effective masses, $P_n$ is the Legendre polynomial of the *n*th order, and $\theta_{eh}$ designates the angle between $r_e$ and $r_h$. The potential energy $V(r_e, r_h)$ describes the motion of the exciton quasi-particle within the NC simulated by an infinitely deep potential well. Thus, $V(r_e, r_h) = 0$ for $r_e, r_h \leq R$ and $V(r_e, r_h) = \infty$ for $r_e, r_h > R$. In Eq. 2, the dielectric-stiffness parameter, $(\bar{\varepsilon}_n)^{-1}$, is defined by $(\bar{\varepsilon}_n)^{-1} = (n+1)(\varepsilon - 1) \cdot \{\varepsilon_r^c(n\varepsilon + n + 1)\}^{-1}$, where $\varepsilon = \varepsilon_r^c/\varepsilon_r^s$. The first two terms in Eq. 2 represent the kinetic energy and the third term depicts the direct Coulomb interaction between an electron and a hole. The two terms that follow the third term correspond to the surface polarization energy which arises from the difference in the dielectric constant ($\varepsilon_r$) between the NC core and the surrounding solvent medium. The former (*i.e,* 4th term) depicts the repulsive self-energy of an electron and a hole due to its own image charge ($V_{ee'}$ and $V_{hh'}$), whereas the latter (*i.e,* 5th term) represents the attractive interaction energy between an electron and a hole *via* image charges ($V_{eh'}$ and $V_{he'}$) (*19*).

The expectation value integral of the above Hamiltonian operator, as calculated using the ground-state (*s-s*)-type wave function of exciton, can be considered as the net exciton energy above the bulk band gap, $\Delta E_g$ (*19,31*). It has two distinctive contributions, repulsive kinetic-energy terms $\langle E_k \rangle$ and Coulomb-like attractive terms $\langle V_c \rangle$ (*31*):

$$\langle \psi_o^s(r_e, r_h; \alpha) | \mathcal{H}_{NC} | \psi_o^s(r_e, r_h; \alpha) \rangle \equiv \Delta E_g(R; \alpha) = \langle E_k \rangle + \langle V_c \rangle$$

$$= \frac{\hbar^2}{2\mu_o}\left\{\frac{\pi^2}{R^2} + \left(\frac{1}{\alpha}\right)^2\right\} - \frac{e^2}{4\pi\varepsilon_o\varepsilon_r^c}\left\{\frac{C_1'}{R} + \frac{C_2}{\alpha}\right\} \geq 0 \tag{3}$$

where $\mu_o$ is the effective reduced mass of exciton $[(\mu_o)^{-1} \equiv (m_e)^{-1} + (m_h)^{-1}]$ and $C_2 = 0.498$. Eq.

3 is valid for $R \leq R_c$, where $R_c$ is the critical size that corresponds to the onset of PL blue shift. The Coulomb interaction coefficient ($C_1'$) can be viewed as the effective Madelung constant (33) for a sphere having radius $R$. Eq. 3 is the most elaborate expression of the net exciton energy above the bulk band gap $[\Delta E_g(R;\alpha) = \langle E_k \rangle + \langle V_c \rangle \geq 0]$, which considers the internal relative motion between the electron and hole in a given exciton (31). This internal correlation effect can be implemented by including the e-h correlation parameter ($\alpha$) in the ground-state wave function $\psi_o^S(r_e, r_h; \alpha)$, where $\psi_o^S(r_e, r_h; \alpha)$ is given by (31).

$$\psi_o^S(r_e, r_h; \alpha) = N j_o\left(\frac{\pi r_e}{R}\right) j_o\left(\frac{\pi r_h}{R}\right) \exp\left(-\frac{r_{eh}}{\alpha}\right) \tag{4}$$

where $j_o$ is the zeroth degree spherical Bessel function and $r_{eh}$ denotes the distance between the electron and hole in each exciton quasi-particle. The normalization constant ($N$) can be evaluated by imposing the normalization condition (31).

To assess the dielectric-confinement effect, we consider the two terms of $\mathcal{H}_{NC}$ that corresponds to the surface polarization energy ($E_{SP}$). The 4th term of Eq. 2 describes to the repulsive self-interaction energy ($\mathcal{H}_{SI} = V_{ee\prime} + V_{hh\prime}$) of the hole and electron with their own images. On the contrary, the 5th term depicts the energy of interaction of the hole and electron with the 'wrong' images ($\mathcal{H}_{WI} = V_{eh\prime} + V_{he\prime}$). According to our notation, the expectation value of this self-polarization term $\langle E_{SP} \rangle$ can be written as (24)

$$\langle E_{SP} \rangle = \langle \psi_o^S(r_e, r_h; \alpha) | \mathcal{H}_{SP} | \psi_o^S(r_e, r_h; \alpha) \rangle = \frac{\{2\pi^{-2} I_{sp} - (\bar{\varepsilon}_0)^{-1}\} e^2}{4\pi \varepsilon_o \varepsilon_r^c R} \equiv \frac{+\Delta C e^2}{4\pi \varepsilon_o \varepsilon_r^c R} \tag{5}$$

where $\mathcal{H}_{SP} = \mathcal{H}_{SI} + \mathcal{H}_{WI}$, $(\bar{\varepsilon}_0)^{-1} = (\varepsilon_r^s)^{-1} - (\varepsilon_r^c)^{-1}$, and the surface-polarization integral ($I_{sp}$) is defined by the following equation (24): $I_{sp} = \pi \int_0^\pi dx \sin^2 x \sum_{n=0}^\infty (\bar{\varepsilon}_n)^{-1} \left(\frac{x}{\pi}\right)^{2n}$. Therefore, the effective Coulomb interaction coefficient ($C_1'$) modified by the dielectric-confinement effect is given by

$$C_1' \equiv C_1 - \Delta C = C_1 - \{2\pi^{-2} I_{sp} - (\bar{\varepsilon}_0)^{-1}\} = -\frac{1}{2} A_1 \tag{6}$$

As shown in Eq. 6, there is a distinct difference between the two contributions, $C_1$ and $\Delta C$, to the modified interaction coefficient. The unmodified Coulomb interaction coefficient $C_1$ (= 1.786) is related exclusively to the pure quantum-confinement effect in the absence of any low-permittivity dielectric medium (31). The presence of the surrounding low-permittivity dielectric medium provides an additional confinement effect on a NC core (24). This repulsive dielectric-confinement effect always decreases the effective Coulomb-interaction coefficient by $\Delta C$. The last expression of Eq. 6 is written to correlate $C_1'$ with the notation of Takagahara (24). Thus, $A_1 = -2C_1'$. $A_1$ was numerically evaluated as a function of the permittivity ratio, $(\varepsilon_r^c / \varepsilon_r^s)$ (24). Accordingly, one can evaluate $C_1'$ for various values of $(\varepsilon_r^c / \varepsilon_r^s)$ using Eq. 6. As reflected in Eq. 6, $I_{sp}$ is a measure of the repulsive dielectric-confinement effect. On the contrary, $(\bar{\varepsilon}_0)^{-1}$ (> 0) tends to increase the degree of attractive

Coulomb interaction and gives a negative contribution to the dielectric-confinement effect.

We examined the role of the internal *e-h* correlation parameter ($\alpha$) in determining the NC size at $R_c$ that corresponds to the onset of PL blue shift. If $\alpha$ were absent, one would expect a continuous PL red shift beyond the critical size $R_c$ until the NC size increases to $2R_c$ which is equal to $R_{eq}$ (Fig. 2B), where $\left(\partial \Delta E_g/\partial R\right)_{\mu_o,\varepsilon_r^c} = 0$ (Supplementary Materials). This is clearly in contradiction to the experimental observations (*12,13,34*) that $R_c$ is the onset of PL blue shift with decreasing NC size or equivalently the surceasing point of PL red shift with increasing NC size beyond this value. This indicates that $\alpha$ plays an important role in determining the NC size corresponding to the onset of PL blue shift. In addition to this finding, we theoretically showed that there exists 1:1 correspondence between the critical size $R_c$ and the corresponding value of the correlation parameter $\alpha_c$. We further showed thermodynamic stability of exciton at $R_c$, which requires (i) $\left(\partial \Delta E_g/\partial R\right)_{\mu_o,\varepsilon_r^c} = 0$ and (ii) $\left(\partial^2 \Delta E_g/\partial R^2\right)_\alpha > 0$ (Fig. 2C). Detailed discussion and theoretical analysis on the relation between $R_c$ and $\alpha_c$ and thermodynamic stability are given in the Supplementary Materials.

**Exciton Bohr radius of spherical NCs under dielectric confinement**

We first consider the ratio of the two proportionality constants appeared in Eq. 3 to derive an expression of the exciton Bohr radius ($a_B$) in terms of $C_1', C_2$, and $R_c$. According to Eq. 1, this ratio is $(e^2/4\pi\varepsilon_o\varepsilon_r^c)/(\hbar^2/2\mu_o) = 2/a_B$. Eq. 3 further indicates that $\Delta E_g(R;\alpha) = 0$ at the critical NC size ($R_c$) for the onset of PL blue shift. Combining Eq. 3 with Eq. 1 under $\Delta E_g(R;\alpha) = 0$ then yields the following relation at $R_c$ under the dielectric confinement:

$$\frac{1}{2}a_B = \frac{(C_1'\alpha_c + C_2 R_c) R_c \alpha_c}{(\pi^2 \alpha_c^2 + R_c^2)} \tag{7}$$

where $\alpha_c$ denotes the *e-h* correlation parameter at the onset size of the PL blue shift, as explained previously. We define the ratio of $\alpha$ to $a_B$ by $\gamma \equiv \alpha/a_B$ for convenience. Combing this definition with Eq. 7 yields a quadratic equation for $R_c$. Solving the resulting quadratic equation yields the following expression of $(R_c/a_B)$ in terms of $C_1', C_2$, and $\gamma_c$:

$$\left(\frac{R_c}{a_B}\right) = \frac{\xi_R' \left\{-1 + \left(1 + \frac{\nu \pi^2}{C_1' \xi_R'}\right)^{1/2}\right\}}{\nu} \tag{8}$$

where $\xi_R' \equiv C_1' \gamma_c^2$ and $\nu \equiv 2\left(C_2 \gamma_c - \frac{1}{2}\right)$. This is a self-consistent equation for $a_B$ because $\gamma_c$ in the right-hand side of Eq. 8 is defined as $\gamma_c \equiv \alpha_c/a_B$. According to the quantum variational calculations (Fig. 3A), there exists one unique value of $\gamma_c$ for a fixed value of $(R_c/a_B)$. Eq. 8 is a generic relationship which is valid for any spherically shaped photoluminescent NC under the dielectric confinement, where the following condition is satisfied: $\varepsilon_r^s < \varepsilon_r^c$. Eq. 8 was obtained by imposing the condition that $\Delta E_g(R; \alpha) = 0$ at $R_c$.

One must find a suitable value of $C_1'$ before solving Eq. 8 self-consistently. Our dispersion medium consists of toluene ($\varepsilon_r = 2.4$) with a minor amount of oleic acid ($\varepsilon_r = 2.5$) as a capping agent (*13*). In view of this, we take $\varepsilon_r^s \approx 2.5$ for the dispersion medium. Then, one can evaluate $C_1'$ for various values of $(\varepsilon_r^c/\varepsilon_r^s)$, as described previously. The self-consistently obtained solution of $a_B$ also should satisfy Eq. 1. We can simultaneously determine $a_B$ and $\varepsilon_r^c$ for a fixed value of $\mu_o$ by imposing this requirement. This is another outstanding point of the present method of assessing $a_B$ without adopting any assumption. We found that the optimized value of $(R_c/a_B)$ for the dispersion of MAPbBr3 perovskite NCs is 3.15 with $C_1'$ equal to 1.31. Thus, the exciton Bohr radius $(a_B)$ under the dielectric confinement is **4.36 nm** in that the onset size of the PL blue shift is 13.75 nm for MAPbBr3 (Fig. 1 and Fig. 4A). This value is within the range of other reported values (4~5 nm) for MAPbBr3 (*13,16-19,35*). The corresponding dielectric constant $(\varepsilon_r \approx \varepsilon_r^c)$ is 9.65, as estimated using Eq. 1 with the reported value of $0.117 m_o$ for $\mu_o$ (*16,21*). This value of $\varepsilon_r$ for MAPbBr3 is slightly smaller than the reported value of 10.75 (*17*) but is bigger than $\varepsilon_r$ of 7.5 as obtained by using magneto-optical absorption spectra (*21*). Considering these two reported values, we judge that our theoretically estimated value of 9.65 seems to be reliable. Computational details of $a_B$ and $\varepsilon_r^c$ are given in **MATERIALS AND METHODS**.

**Exciton Bohr radius without dielectric confinement**

We then consider a colloidal dispersion that would be characterized by the absence of any dielectric confinement. In this hypothetical dispersion, the dielectric constant of the dispersed NCs is equal to that of the dispersion medium, *i.e.*, $\varepsilon_r^c = \varepsilon_r^s$. Under this condition, $\Delta C = 0$ in Eq. 6. In other words, the effective Coulomb interaction coefficient $(C_1')$ would be equal to $C_1 (= 1.786)$. Then, $\Delta E_g(R; \alpha) = 0$ (Eq. 3) with $C_1'$ replaced by $C_1$ eventually yields the following $(R_c/a_B)$ ratio in the absence of dielectric confinement:

$$\left(\frac{R_c}{a_B}\right) = \frac{\xi_R\left\{-1 + \left(1 + \frac{\nu \pi^2}{C_1 \xi_R}\right)^{1/2}\right\}}{\nu} \qquad (9)$$

where $\xi_R \equiv C_1 \gamma_c^2$ and $\nu \equiv 2\left(C_2 \gamma_c - \frac{1}{2}\right)$. Notice that $\xi_R$ is now defined in terms of $C_1 (= 1.786)$, not in terms of $C_1'$. As explained previously, there exists one unique value of $\gamma_c (\equiv \alpha_c/a_B)$ for fixed values of $(R_c/a_B)$, $C_1$, and $C_2$. Then, we can readily show the following universal solution of $(R_c/a_B)$, which would be valid for any colloidal NCs in the absence of the dielectric-confinement effect:

$$\left(\frac{R_c}{a_B}\right) = 2.45 \qquad (10)$$

We find that the corresponding $\gamma_c$ is equal to 1.36 (Fig. 3A). The $(R_c/a_B)$ ratio is significantly reduced from the corresponding value under the dielectric confinement (3.15 in Eq. 8). Accordingly, the exciton Bohr radius is expected to increase substantially upon removing the dielectric-confinement effect. Substituting $R_c = 13.75\ nm$ into Eq. 10 immediately yields 5.61 nm for $a_B$ with the corresponding dielectric constant ($\varepsilon_r^C$) of 12.41 (See **MATERIALS AND METHODS**).

According to Eq. 1, the enhanced $a_B$, from 4.36 to 5.61 $nm$, is a direct consequence of the increased $\varepsilon_r^C$ or *vice versa* upon removal of the dielectric-confinement effect. In the absence of the dielectric-confinement effect, the quantum *e-h* correlation effect also controls the magnitude of $a_B$. We arrive at this conclusion by considering the variation of $(R_c/a_B)$ value upon removal of the quantum *e-h* correlation effect in the absence of dielectric-confinement effect where $C_1' = C_1$. Under this condition, we can readily deduce that $(R_c/a_B) = \pi^2/2C_1 \approx 2.75$ by combining Eq. 3 with Eq. 1, as we predicted previously (*26*). Comparing this hypothetical case of $(R_c/a_B) \approx 2.75$, where $\alpha \to \infty$, with Eq. 10 suggests that the quantum *e-h* correlation effect (with $\alpha = 1.36$) tends to increase $a_B$ significantly. In the next section, we will show that there is a good linear correlation between $a_B$ and the quantum *e-h* correlation energy over a wide range of $\varepsilon_r^S$ values.

It should be emphasized that Eq. 9 is not valid for any real colloidal dispersion in which $\varepsilon$ ($\equiv (\varepsilon_r^C/\varepsilon_r^S)$) is bigger than 1 for the dispersion stability. In the case of MAPbBr$_3$ colloidal NCs, $\varepsilon = 9.65/2.5 = 3.86 \gg 1$. Thus, Eq. 8, instead of Eq. 9, with $C_1'$ ($< C_1$) should be used to estimate $a_B$ for colloidally dispersed NCs under the dielectric confinement. At this stage, we have the following intriguing question: Is there any real system that is represented by Eq. 9 for deducing $a_B$ in the absence of dielectric confinement? A macroscopic-scale photoluminescent single crystal can be a suitable candidate for fulfilling this requirement. It is because a single crystal can be viewed as a hypothetical colloidal dispersion in which the NC cores having the characteristic $R_c$ are dispersed in a dielectric medium with its dielectric constant $\varepsilon_r^S (= \varepsilon_r^C)$ equal to that of the NC cores (Fig. 2D). Here, $R_c$ is a material parameter and thus has a characteristic value for a given material: $R_c = 13.75\ nm$ for MAPbBr$_3$. The dielectric-confinement effect disappears in this single crystal owing to the absence of dielectric boundary between the imaginary core with the radius $R_c$ and the dielectric medium which is factually a single-crystalline matrix (Fig. 2D). Thus, the effective Coulomb interaction coefficient $C_1'$ is equal to $C_1$ ($= 1.786$). Because of this equality, $a_B$ of the macroscopic-scale single crystal can be evaluated by applying Eq. 9 or 10 with $C_1 = 1.786$. Accordingly, we conclude that $a_B$ of MAPbBr$_3$ single crystal is 5.61 nm with $\varepsilon_r^C$ of 12.41 (Table 1). Practically, this argument can be extended to a dense polycrystalline film without having a low permittivity grain-boundary phase because the average radius of grains ($R_g$) is much bigger than $R_c$ (= 13.75 nm for MAPbBr$_3$) in general. Therefore, $a_B$ of a polycrystalline MAPbBr$_3$ film having large composing grains is expected to be about 5.61 nm.

**Effect of the degree of dielectric confinement on the exciton Bohr radius**

We have self-consistently estimated important material parameters as a function of $\varepsilon_r^s$ to assess the effect of the degree of dielectric confinement on the variation of these parameters that include $a_B$ and $\varepsilon_r^c$ (Table 2). The effective Coulomb interaction coefficient ($C_1'$) increases as $\varepsilon_r^s$ increases (Fig. 5A). $C_1'$ reaches its maximum value of 1.786 when the ($\varepsilon_r^c/\varepsilon_r^s$) ratio is 1.0. This is because $C_1'$ becomes $C_1$ (= 1.786) when the dielectric-confinement effect disappears at $\varepsilon_r^s = \varepsilon_r^c = 12.41$ for MAPbBr$_3$. Similar to $C_1'$, the dielectric constant of the NC core ($\varepsilon_r^c$) increases as $\varepsilon_r^s$ increases (Fig. 5A). The exciton Bohr radius also tends to increase with increasing value of $\varepsilon_r^s$ (Fig. 5B), reaching its maximum value of 5.61 nm when the ($\varepsilon_r^c/\varepsilon_r^s$)-ratio is 1.0 at $\varepsilon_r^s = 12.41$ (Table 1). In other words, $a_B$ tends to decrease with decreasing value of $\varepsilon_r^s$ and this reducing tendency becomes progressively important with the degree of dielectric confinement (Fig. 5B).

According to Eq. 3, two distinctive factors have a significant influence on the degree of Coulomb-like attraction between electron and hole within an exciton pair: (i) the effective classical Coulomb attraction within an exciton pair as described by $-e^2 C_1'/4\pi\varepsilon_o\varepsilon_r^c R_c$ at $R_c$ and (ii) the attractive quantum correlation effect as measured by the parameter $\alpha$. Thus, the classical Coulomb attraction is proportional to $C_1'/\varepsilon_r^c$-ratio at a fixed $R_c$. We have computed this ratio for a wide variety of $\varepsilon_r^s$ values. Surprisingly, the $C_1'/\varepsilon_r^c$-ratio is essentially constant with a characteristic value of 0.14 over a wide range of $\varepsilon_r^s$ between 1.5 and 12.41. This suggests that the classical Coulomb attraction energy is essentially independent of the degree of dielectric confinement, *i.e.*, $\varepsilon_r^s$. Accordingly, $a_B$ would remain at an essentially constant value if the classical Coulomb attraction were primarily responsible for determining the magnitude of $a_B$ over a wide range of $\varepsilon_r^s$. Contrary to this expectation, the estimated $a_B$ does not remain at a constant value but tends to decrease as $\varepsilon_r^s$ decreases. This reducing $a_B$ is especially pronounced at small $\varepsilon_r^s$ values (Fig. 5B).

The above semi-empirical finding suggests that the dielectric-confinement effect on $a_B$, thus, $\varepsilon_r^s$-dependent $a_B$, is primarily due to the Coulomb-like quantum correlation effect as quantified by the quantum *e-h* correlation parameter $\alpha$. The quantum *e-h* correlation energy ($E_{corr}$) at $R_c$ can be written, from Eq. 3, as

$$E_{corr} = \frac{-e^2 C_2}{4\pi\,\varepsilon_o \varepsilon_r^c \alpha_c} = \frac{-e^2 C_2}{4\pi\,\varepsilon_o(\varepsilon_r^c \gamma_c a_B)} \tag{11}$$

It is notable that $\alpha_c$ (= $\gamma_c a_B$) has a dimension of length. If we replaced $\alpha_c$ with $R_c$ and $C_2$ with $C_1'$, Eq. 11 would represent the classical Coulomb attraction between two opposite charges separated by the distance of $R_c$. This is the reason why we name 'Coulomb-like' interaction for $E_{corr}$. Thus, $|E_{corr}|$ is proportional to $(\varepsilon_r^c \gamma_c a_B)^{-1}$. Considering this proportionality, we have plotted the estimated $a_B$ values as a function of $(\varepsilon_r^c \gamma_c a_B)^{-1}$. As shown in Fig. 5C, there is a good linear correlation between $a_B$ and $(\varepsilon_r^c \gamma_c a_B)^{-1}$ with the slope being $-133\,(nm)^2$. We then extract the following numerical linear relation using materials parameters listed in Table 2: $a_B(nm) = -133\,(nm)^2 \cdot (\varepsilon_r^c \gamma_c a_B)^{-1} + 7.08\,(nm)$. Combining this linear relation with Eq. 11 yields a semi-empirical linear correlation between $a_B$ and $|E_{corr}|$: $a_B(nm) = -532\,\pi\varepsilon_o(e^2 C_2)^{-1} |E_{corr}| + 7.08\,(nm)$. This semi-empirically deduced relation shows that there is a linear correlation between $a_B$ and $|E_{corr}|$ with the slope being $= -532\pi\varepsilon_o(e^2 C_2)^{-1}$. Therefore, we can conclude that the attractive quantum *e-h* correlation has a direct influence on controlling the magnitude of $a_B$ over a wide range of $\varepsilon_r^s$ between 1.5 and 12.41 (*i.e.*, over a wide range of the degree of dielectric confinement).

**Exciton Bohr radius of cubic-shaped NCs**

The starting point of theoretical study on $a_B$ is the net exciton energy above the bulk band gap $\Delta E_g(L; \alpha)$, which is given by the sum of the expectation value of the exciton's kinetic energy $\langle E_k \rangle$ and that of the Coulomb-like attraction $\langle V_c \rangle$ of a cubic-shaped NC. We obtain the following expectation value integral of the kinetic energy for an exciton confined in a cube with the edge length $L$:

$$\langle E_k \rangle = \langle \psi_o(r_e, r_h; \alpha) | \tilde{T} | \psi_o(r_e, r_h; \alpha) \rangle = \frac{\hbar^2}{2\mu_o}\left\{\frac{3\pi^2}{L^2} + \left(\frac{1}{\alpha}\right)^2\right\} \tag{12}$$

where $\tilde{T}$ denotes the kinetic energy operator and $\psi_o(r_e, r_h; \alpha)$ designates the ground-state exciton wave function for a cube with the edge length $L$. It is given by $\psi_o(r_e, r_h; \alpha) = \phi_o(r_e, r_h) \cdot \chi(\alpha) = N \cos(kx_e)\cos(kx_h) \cdot \cos(ky_e)\cos(ky_h) \cdot \cos(kz_e)\cos(kz_h) \cdot \exp(-r_{eh}/\alpha)$, where $k = \pi/L$ (see the Supplementary Materials for detailed theoretical derivations).

On the contrary, the expectation value integral of the Coulomb-like attractive interaction is given by the ratio of two 3-fold integrals (*36*)

$$\langle V_c \rangle = -\frac{e^2}{4\pi\varepsilon_o \varepsilon_r^c}\left(\frac{\pi}{L}\right)\frac{I_{-1}(\beta)}{I_0(\beta)} \approx -\frac{e^2}{4\pi\varepsilon_o \varepsilon_r^c}\left\{\frac{C_1'}{R_{eff}} + \frac{C_2}{\alpha}\right\} \tag{13}$$

where $\beta = k\alpha = \pi\alpha/L$ and $I_n(\beta)$ is given by the following-type 3-fold integral with respect to $\xi_x, \xi_y, \xi_z$: $I_n(\beta) = 2^{-6}\iiint_0^\pi d\xi_x d\xi_y d\xi_z\, g(\xi_x)g(\xi_y)g(\xi_z) \exp\left(-\frac{2}{\beta}\sqrt{\xi_x^2 + \xi_y^2 + \xi_z^2}\right)\left(\xi_x^2 + \xi_y^2 + \xi_z^2\right)^{n/2}$ with $\xi_x = \xi_{ex} - \xi_{hx} = k(x_e - x_h)$, *etc.* Here, $g(\xi)$ is defined by $g(\xi) = (\pi - \xi)\{2 + \cos(2\xi)\} + \frac{3}{2}\sin(2\xi)$. Thus, we cannot obtain an analytical expression of $\langle V_c \rangle$ in terms of $L$ (edge length) and $\alpha$. Considering this mathematical limitation, we are seeking a suitable analytical expression of $\langle V_c \rangle$ which is similar to $\langle V_c \rangle$ for spherical NCs. To this end, we first consider the effective Madelung constant ($C_1$) which is a measure of the *e-h* Coulomb attraction within a cubic-shaped NC. It is known that the effective Madelung constant of a monoatomic crystal with the body-centered cubic (bcc) or face-centered cubic (fcc) symmetry is 1.79 (*37*). This value is nearly equal to that of a spherical NC in the absence of dielectric confinement, *i.e.*, $C_1 = 1.786$. Considering this, we judge that the effective

Madelung constant practically remains invariant when a NC changes its morphological symmetry from spherical to cubic. Therefore, $\langle V_c \rangle$ of a cubic-shaped NC with the 'hypothetical' effective radius $R_{eff}$ can be written using the right-hand side expression of Eq. 13. We are now ready to obtain an analytical expression of the net exciton energy above the bulk band gap, $\Delta E_g(L;\alpha)$, using Eqs. 12 and 13 with the equality that $L^3 = (4\pi/3)R_{eff}^3$.

$$\Delta E_g(L;\alpha) = \langle E_k \rangle + \langle V_c \rangle = \frac{\hbar^2}{2\mu_o}\left\{\frac{3\pi^2}{L^2} + \left(\frac{1}{\alpha}\right)^2\right\} - \frac{e^2}{4\pi\varepsilon_o \varepsilon_r^c}\left\{\left(\frac{4\pi}{3}\right)^{1/3}\frac{C_1'}{L} + \frac{C_2}{\alpha}\right\} \quad (14)$$

Then, combining Eq. 1 with Eq. 14 under $\Delta E_g(L;\alpha) = 0$ yields the following relation at the onset size ($L_c$) of the PL blue shift:

$$\frac{1}{2}a_B = \frac{\{1.612\, C_1'\alpha_c + C_2 L_c\} L_c \alpha_c}{3\pi^2 \alpha_c^2 + L_c^2} \quad (15)$$

Again, we define $\alpha_c$ as $\gamma_c a_B$. Combing this definition of $\alpha_c$ with Eq. 15 yields a quadratic equation for $L_c$. Solving the resulting quadratic equation yields the following expression of $(L_c/a_B)$ in terms of $C_1', C_2,$ and $\gamma_c$:

$$\left(\frac{L_c}{a_B}\right) = \frac{\left(-1.612 C_1'\gamma_c^2 + \gamma_c\sqrt{(1.612)^2(C_1')^2\gamma_c^2 + 6\left(C_2\gamma_c - \frac{1}{2}\right)\pi^2}\right)}{2\left(C_2\gamma_c - \frac{1}{2}\right)} = \frac{\xi_L'\left\{-1 + \left(1 + \frac{3\nu\pi^2}{1.612\, C_1'\xi_L'}\right)^{1/2}\right\}}{\nu} \quad (16)$$

where $\xi_L' \equiv 1.612 C_1'\gamma_c^2$ and $\nu \equiv 2\left(C_2\gamma_c - \frac{1}{2}\right)$. Eq. 16 is a generic relationship which is valid for any cubic-shaped photoluminescent NC under the dielectric confinement, where the following condition is satisfied: $\varepsilon_r^s < \varepsilon_r^c$.

Taking $\varepsilon_r^s = 2.5$, we evaluate $C_1'$ for various values of $(\varepsilon_r^c/\varepsilon_r^s)$, as explained previously. The self-consistently obtained solution of $a_B$ also should satisfy Eq. 1. Taking the reported value of $\mu_o = 0.07 m_o$ for CsPbBr$_3$ (7) and using Eqs. 1 and 16, we can simultaneously determine $a_B$ and $\varepsilon_r^c$ without adopting any assumption. We found that the optimized value of $(L_c/2a_B)$ for the cubic-shaped CsPbBr$_3$ NCs dispersion is 2.66 with $C_1' = 1.45$ and $\gamma_c = 1.28$. Thus, the exciton Bohr radius ($a_B$) under the dielectric confinement is **5.08 nm** in that the onset size ($L_c$) of the PL blue shift for CsPbBr$_3$ is 27.0 nm (Fig. 4B). The corresponding dielectric constant ($\varepsilon_r^c$) is 6.72, as evaluated using Eq. 1 with $\mu_o = 0.07 m_o$.

There appears a noticeable difference in $a_B$ between the present theoretical value (5.08 nm) and the reported value of 3.5 nm (7). This disagreement is mainly originated from the difference in the dielectric constant ($\varepsilon_r^c$) between the two studies in that both studies adopt the same $\mu_o$ value ($0.07 m_o$). The present value of $\varepsilon_r^c$ (6.72) is 1.35 times bigger than the DFT computed high-frequency limit value of 4.96 (7). This ratio, *i.e.*, 1.35, is close to the $a_B$ ratio between the two studies, namely, 5.08/3.5 = 1.45. We further predict that $\varepsilon_r^c$ of CsPbBr$_3$ single crystal is as high as 7.94 in the absence of the dielectric-confinement effect. The reported dielectric constants of various lead halide perovskites-based polycrystalline films are between 7.5 and 11.9 based on careful data fitting of magneto-optical absorption spectra (*21*). This suggests that $\varepsilon_r^c$ values obtained by the present self-consistent calculations (6.72~7.94) is more appropriate than the computed high-frequency limit value (4.96) in describing the screening effect on the effective Coulomb interaction between electron and hole in a given exciton pair.

The exciton Bohr radius ($a_B$) in the absence of the dielectric-confinement effect can be obtained by using Eq. 16 with the replacement of $C_1'$ by $C_1$ (= 1.786). The self-consistently obtained ($L_c/2a_B$) value of the cubic CsPbBr$_3$ in the absence of the dielectric-confinement effect is 2.25 with $\gamma_c = 1.33$. Thus, $a_B$ of CsPbBr$_3$ single crystal is **6.00 nm** as $L_c$ is 27.0 nm (Fig. 4B). The corresponding dielectric constant ($\varepsilon_r^c$), as computed using Eq. 1 with $\mu_o = 0.07 m_o$, is 7.94 which is ~18% increase as compared with $\varepsilon_r^c$ under the dielectric confinement (6.72). Thus, the enhanced $a_B$ value (from 5.08 to 6.00 nm) upon removing the dielectric-confinement effect is mainly attributed to the enhanced screening in the Coulomb interaction (*i.e.*, $\varepsilon_r^c$) between electron and hole in each exciton pair. The self-consistently calculated values of $a_B, \varepsilon_r^c$, and $\gamma_c$ for both MAPbBr$_3$ and CsPbBr$_3$ are summarized in Table 1.

**Exciton binding energy with and without dielectric confinement**

The ratio of the radiative recombination of *e-h* pairs to the separation into free-charge carriers is highly important both in LED devices and photovoltaic cells. It is defined as $R_{REC}$. Thus, $R_{REC} \equiv n_{exc}/n_{FC} = (n - n_{FC})/n_{FC}$, where $n_{exc}$ is the density of *e-h* exciton pairs, $n_{FC}$ is the density of free-charge carriers, and $n$ is the total density of excitation (= $n_{FC} + n_{exc}$). In metal halide perovskites (MHPs), $R_{REC}$ can be estimated by applying the Saha-Langmuir equation (*38*).

$$\frac{n \cdot x^2}{1-x} = \frac{n_{FC}^2}{n_{exc}} = \left(\frac{2\pi \mu_o k_B T}{h^2}\right)^{3/2} e^{-E_b/k_B T} \qquad (17)$$

where $x$ denotes the fraction of free carriers over the total density (= $n_{FC}/n$) and $E_b$ is the exciton binding energy that signifies the energy needed to separate the bound *e-h* pairs. Numerical estimates based on the Saha-Langmuir equation indicated that, for a fixed total density of excitation ($n$), $1/x$ increases as $E_b$ increases (*26,38*). It can be shown immediately that $R_{REC} + 1 = 1/x$. Therefore, $R_{REC}$ is determined by $E_b$. Consequently, the rate of radiative recombination of *e-h* pairs is expected to increase as $E_b$ increases.

We have estimated $E_b$ under the dielectric confinement by evaluating the Coulomb-like attractive interaction terms. Since $E_b$ is the energy needed to separate the bound *e-h* pairs, we write the following effective relation for $E_b$ of colloidally dispersed NCs at $R_c$ (*i.e.*, under the dielectric confinement) by considering the two relevant attractive terms in Eq. 3:

$$E_b(C_1') = \frac{e^2}{4\pi\varepsilon_o\varepsilon_r^c}\left\{\frac{C_1'}{R_c}+\frac{C_2}{\alpha_c}\right\} = \frac{\hbar^2}{a_B\,\mu_o}\left\{\frac{C_1'}{R_c}+\frac{C_2}{\gamma_c a_B}\right\} = 149.2\ (meV) * (0.095 + 0.090) = 27.6\ (meV) \qquad (18)$$

We used Eq. 1 and $\alpha_c = \gamma_c a_B$ in obtaining the third expression of Eq. 18. We obtained numerical value of $E_b$ by plugging the following previously optimized values for colloidally dispersed MAPbBr$_3$ NCs at $R_c$: $a_B = 4.36 \times 10^{-9}(m)$, $\gamma_c = 1.27$, $C_1' = 1.31$. In addition, we used the following reported values: $C_2 = 0.498$ *(31)*, $\mu_o = 0.117 m_o$ *(16,21)*, and $R_c = 13.75 \times 10^{-9}(m)$ as a characteristic material parameter. It is interesting to note that the quantum correlation effect ($C_2/\alpha_c$) and the classical Coulombic attraction ($C_1'/R_c$) almost equally contribute to $E_b$, namely, $0.090: 0.095 \approx 1: 1$. Thus, $E_b$ for a colloidally dispersed MAPbBr$_3$ is 27.6 *meV* under the dielectric confinement.

On the other hand, $E_b$ in the absence of dielectric confinement can be obtained by plugging the following previously optimized values for a single crystalline MAPbBr$_3$ where $\varepsilon_r^c = \varepsilon_r^s = 12.41$ (Table 2): $a_B = 5.61 \times 10^{-9}(m)$, $\gamma_c = 1.36$, $C_1' = C_1 = 1.786$, and $C_2 = 0.498$. Thus,

$$E_b(C_1) = \frac{e^2}{4\pi\varepsilon_o\varepsilon_r^c}\left\{\frac{C_1}{R_c}+\frac{C_2}{\alpha_c}\right\} = \frac{\hbar^2}{a_B\mu_o}\left\{\frac{C_1}{R_c}+\frac{C_2}{\gamma_c a_B}\right\} = 116.0\ (meV) * (0.130 + 0.065) = 22.6\ (meV) \qquad (19)$$

Comparing this value with that of colloidally dispersed NCs indicates that the dielectric-confinement effect tends to increase $E_b$ significantly, which is mainly due to the reduced $a_B$ under the dielectric confinement (4.36 nm *vs.* 5.61 nm). It is interesting to note that this value for a single crystal, 22.6 *meV*, is very close to the reported value of 25 *meV* for a coarse-grained polycrystalline MAPbBr$_3$ film *(21)*, which was obtained by careful data fitting of the magneto-optical absorption spectroscopy.

The exciton binding energy of colloidally dispersed cubic-shaped NCs at $L_c$ can be estimated by considering the relevant attractive terms in Eq. 14. Thus, we obtain.

$$E_b(C_1') = \frac{\hbar^2}{a_B\,\mu_o}\left\{\left(\frac{4\pi}{3}\right)^{1/3}\frac{C_1'}{L_c}+\frac{C_2}{\gamma_c a_B}\right\} = 214.0\ (meV) * (0.087 + 0.077) = 35.1\ (meV) \qquad (20)$$

It is interesting to note again that the quantum correlation effect ($C_2/\alpha_c$) is as important as the classical Coulombic attraction ($C_1'/L_c$) with the relative contribution given by $0.077/0.087 \approx 0.9/1.0$. We obtained numerical value of $E_b$ by plugging the following previously optimized values for colloidally dispersed CsPbBr$_3$ NCs: $a_B = 5.08 \times 10^{-9}(m)$, $\gamma_c = 1.28$, $C_1' = 1.45$, $L_c = 27.0\ (nm)$. In addition, we used the following reported values: $C_2 = 0.498$ *(31)* and $\mu_o = 0.07 m_o$ *(7)*. This value is close to the reported value of 40.0 *meV* for cubic CsPbBr$_3$ NCs *(7)*. On the contrary, $E_b$ in the absence of dielectric confinement can be obtained by plugging the following previously optimized values for a single crystalline CsPbBr$_3$: $a_B = 6.00 \times 10^{-9}(m)$, $\gamma_c = 1.33$, $C_1 = 1.786$. Thus,

$$E_b(C_1) = \frac{\hbar^2}{a_B\,\mu_o}\left\{\left(\frac{4\pi}{3}\right)^{1/3}\frac{C_1}{L_c}+\frac{C_2}{\gamma_c a_B}\right\} = 181.4\ (meV) * (0.107 + 0.062) = 30.7\ (meV) \qquad (21)$$

Similar to the spherically shaped MAPbBr$_3$, the dielectric-confinement effect tends to increase $E_b$ of the cubic-shaped CsPbBr$_3$ significantly.

## DISCUSSION

**Comparison of the present method with a commonly used method for $E_b$**

We will present a commonly used method of estimating $E_b$ and compare the result with our new method as presented in the previous section (**RESULTS**). For this purpose, we first consider the electronic energy in the ground-state *1s* orbital of a hydrogen atom ($E_{1s}$) which is described by the well-known Bohr formula, $E_{1s} = -(m_e/2\hbar^2) \cdot (e^2/4\pi\varepsilon_o)^2 \equiv -R_H$, where $R_H$ denotes the Rydberg constant having a numerical value of 13.6 $meV$. If we replace $m_e$ with $\mu_o$ (reduced mass of exciton), $e^2/1$ with $e^2/\varepsilon_r^c$, we obtain the following expression of $E_b$ in terms of $R_H$ (*25,26*):

$$E_b = R_H \frac{\mu_o}{m_e(\varepsilon_r^c)^2} = \frac{e^2}{8\pi\varepsilon_o \varepsilon_r^c a_B} \tag{22}$$

where the 2<sup>nd</sup> expression of $E_b$ was obtained by incorporating Eq. 1 into the 1<sup>st</sup> expression of Eq. 22. Eq. 22 is widely used in defining $E_b$ in terms of $R_H$ (*17,21,25,26,39*) and can be viewed as a modified Bohr formula for $E_b$.

We calculate $E_b$ of the colloidally dispersed MAPbBr₃ NCs under the dielectric confinement by substituting $a_B = 4.36 \times 10^{-9} (m)$ and $\varepsilon_r^c = 9.65$ (Table 1) into Eq. 22. We obtain $E_b = 17.1\ meV$, which is noticeably smaller than our prediction based on Eq. 18 (27.6 $meV$). The main cause of this substantially underestimated $E_b$ can be readily found by examining the last expression of Eq. 22, namely, $E_b = e^2/4\pi\varepsilon_o\varepsilon_r^c (2a_B)$. Thus, Eq. 22 merely represents the static Coulomb attraction energy between electron and hole with their separation distance equal to the exciton Bohr diameter, $2a_B$, thus neglecting the quantum *e-h* correlation effect which can be as large as 13.4 meV according to Eq, 18 [= $27.6 \times (0.090/0.185)\ meV$]. On the other hand, we calculated $E_b$ of a single crystalline MAPbBr₃ in the absence of the dielectric confinement by substituting $a_B = 5.61 \times 10^{-9} (m)$ and $\varepsilon_r^c = 12.41$ (Table 1) into Eq. 22. The computed $E_b$ is 10.3 $meV$ and is less than 1/2 of our prediction in the absence of the dielectric confinement (22.6 $meV$). Moreover, this value is only 41 % of the reported value of 25 $meV$ for a coarse-grained polycrystalline MAPbBr₃ film (*21*). Thus, $E_b$ based on the modified Bohr formula (*i.e.*, Eq. 22) substantially underestimates the correct value of $E_b$ by completely neglecting the quantum *e-h* correlation effect. This demonstrates the improvement of the present method by incorporating the quantum *e-h* correlation term in theoretical formulations of $E_b$, Eqs. 18 and 19.

**Reduced values of $E_b$ in organic iodide perovskites, RPbI₃**

It is known that $E_b$ of organic bromide perovskites is substantially bigger than that of organic iodide counterparts (*21*): (i) $E_b(\text{MAPbBr}_3) = 25\ meV$ *vs.* $E_b(\text{MAPbI}_3) = 16\ meV$, (ii) $E_b(\text{FAPbBr}_3) = 22\ meV$ *vs.* $E_b(\text{FAPbI}_3) = 14\ meV$, where FA denotes formamidinium [CH(NH₂)₂] group. We obtain the following equation for the ratio of two $E_b$s using Eq. 19 which is relevant to single crystals or coarse-grained polycrystalline films:

$$\frac{E_b(\text{RPbI}_3)}{E_b(\text{RPbBr}_3)} = \frac{\varepsilon_r^c(Br)}{\varepsilon_r^c(I)} \left(\frac{f_I}{f_{Br}}\right) \equiv \frac{\varepsilon_r^c(Br)}{\varepsilon_r^c(I)} (1-\theta) \approx \frac{\varepsilon_r^c(Br)}{\varepsilon_r^c(I)} \qquad (23)$$

where R denotes MA or FA cation and $\varepsilon_r^c(I)$, for example, stands for the dielectric constant of RPbI$_3$ in the absence of dielectric confinement. In Eq. 23, $f_I$ is defined as $\{(C_1/R_{c(I)}) + (C_2/\alpha_{c(I)})\}$, where $R_{c(I)}$ denotes the critical size for the onset of PL blue shift of RPbI$_3$ iodide perovskite. Thus, $R_{c(Br)} = 13.75\ nm$ if R is equal to MA (CH$_3$NH$_3$). In Eq. (23), the $(f_I/f_{Br})$-ratio is defined as $1-\theta$. According to this definition, $\theta$ is 0 if $(f_I/f_{Br}) = 1.0$, where the ratio of two $E_b$s is simply described by the ratio of two distinct dielectric constants, namely, $\varepsilon_r^c(Br)/\varepsilon_r^c(I)$. Thus, $\theta$ is a measure of the deviation from the simplification that $E_b$ of a given organic perovskite is determined simply by its reciprocal dielectric constant, i.e., $1/\varepsilon_r^c$, called dielectric stiffness (*40*).

We are able to estimate the parameter $\theta$ for two typical organic perovskites by using the above $E_b$ values and the following dielectric constants (*21*): (i) $\varepsilon_r^c(\text{MAPbBr}_3) = 7.5$ *vs.* $\varepsilon_r^c(\text{MAPbI}_3) = 9.4$, (ii) $\varepsilon_r^c(\text{FAPbBr}_3) = 8.42$ *vs.* $\varepsilon_r^c(\text{FAPbI}_3) = 9.35$ for the orthorhombic phase. We obtain that $\theta = 0.20$ for $E_b(\text{MAPbI}_3)/E_b(\text{MAPbBr}_3)$ and $\theta = 0.29$ for $E_b(FAPbI_3)/E_b(\text{FAPbBr}_3)$. Therefore, the $E_b$ ratio of MA-based organic perovskites is more strongly influenced by their dielectric constants than that of FA-based organic perovskites. According to Eq. 23 and the estimated values of $\theta$ for two cases, $E_b(\text{RPbI}_3)/E_b(\text{RPbBr}_3)$ should be less than 1. Interestingly, this ratio is independent of the organic group, R, and is equal to 0.64 for both MA and FA. This result accords with the observed difference in $E_b$ between organic bromide perovskites and organic iodide perovskites (*15,21,22*). Therefore, organic iodide perovskites (*e.g.*, MAPbI$_3$ and FAPbI$_3$) are more appropriate than bromide counterparts for photovoltaic applications that require smaller $E_b$ for efficient charge-carriers transport.

## MATERIALS AND METHODS
### Synthesis of perovskite nanocrystals

MAPbBr$_3$ NCs were fabricated using ligand-assisted reprecipitation (LARP) in air at room temperature (*13*). The precursor solution that contained amine ligands was prepared by dissolving 0.3 *mmol* of CH$_3$NH$_3$Br (Dyesol), 0.4 *mmol* of PbBr$_2$ (Aldrich, 99.999%) and 40 *μL* of *n*-hexylamine (Aldrich, 99%) in 10 mL of anhydrous *N, N*-dimethylformamide (DMF, Aldrich). To synthesize medium-sized particles, reagents for precursor and *n*-hexylamine were dissolved in 5 *ml* of DMF in the same molar ratio used to synthesize the NCs. Then 10% of the precursor-mixture volume was injected into 5 *mL* of toluene that included various concentrations of oleic acid (Aldrich, 99%; from 0.5 to 100 *μL*) under vigorous stirring to induce crystallization. After 10 min, the colloidal solutions were centrifuged at 3000 rpm for 10 min to precipitate large particles. The supernatant was collected, and the precipitate was discarded.

Hot-injection method was used for the synthesis of uniformly-sized cubic-shaped CsPbBr$_3$ NCs following the previous report with modifications (*7*). In a typical synthesis procedure, 0.1 mmol of Cs$_2$CO$_3$ (Sigma Aldrich, 99.9%) and 0.2 mmol of PbO (Sigma Aldrich, 99.0%) were combined with 10 ml of 1-octadecene (ODE, Sigma Aldrich, 90%) and 1 ml of oleic acid (OA, Sigma Aldrich, 90%) in a 50 ml three-necked reaction flask. The mixture was stirred under vacuum at 120°C for 30 minutes to ensure complete dissolution of the precursors. Subsequently, the temperature was increased to the desired range (60-280°C) under a continuous flow of N$_2$. An optimized amount of 1.8 ml of OA-HBr was then injected into the flask. After a reaction time of 15 seconds, the flask was immediately quenched in an ice bath. Once the reaction mixture had cooled to room temperature, 20 ml of hexane was added to the reaction flask. The nanocrystals were precipitated by centrifugation at 6000 rpm for 5 minutes. The supernatant was discarded, and the precipitated nanocrystals were re-dispersed and harvested in hexane. The size of CsPbBr$_3$ NCs are conveniently tuned by the reaction temperature.

**Photoluminescence and transmission electron microscopy measurements**

PL spectra of PNCs were acquired using a JASCO FP8500 spectrofluorometer in colloidal dispersion states. The measurements were performed using a 150-W Xenon lamp light source and a PMT detector with excitation wavelength of 405 nm for MAPbBr$_3$ and CsPbBr$_3$ PNCs. A highly concentrated colloidal dispersion is accompanied by a red shift in the PL spectrum owing to the photon reabsorption (*41*). Therefore, the PL measurement was conducted in a highly diluted dispersion that did not show any blue shift upon further dilution (Absorbance < 0.5 at 405 nm). Nanoparticle colloidal dispersions were dropped onto a formvar/carbon supported copper grids immediately after the PL measurement. Transmission electron microscopy (TEM) measurement was conducted using a Tecnai F20 at an acceleration voltage of 200 kV.

**Image cytometry analysis**

The size information for each spherically shaped particle was extracted using image-analysis software (ImageJ 1.41n, NIH, USA) except small particles (< 5 nm) due to a low contrast ratio. The obtained bright-field optical TEM images were converted to 8-bit images after adjusting the contrast ratio and brightness. Then the areas were acquired and used to calculate the diameter (*12*). More than 400 particles were measured at each nominal size for MAPbBr$_3$. In the case of CsPbBr$_3$, more than 100 particles were measured at each nominal size. The small particles (< 5 nm) of MAPbBr$_3$ were close to perfectly circular and uniform, so the diameter of a small particle was directly measured by marking opposite ends. For this measurement, the modal diameter, which corresponds to the peak in the size-distribution histogram, was obtained by measuring more than 200 particles in each.

**Self-consistent iterative calculations of $a_B$ and $\varepsilon_r^c$**

We first assign the input value of $(R_c/a_B)$ to accurately evaluate $a_B$ and $\varepsilon_r^c$ by performing self-consistent iterative calculations. Assume that $(R_c/a_B) = 3.0$ as the 1st input value for colloidal MAPbBr$_3$ under the dielectric confinement. Then, the 1st input value of $a_B$ is 4.58 nm as $R_c = 13.75 \, nm$. We obtain $\gamma_c = 1.28$ for $(R_c/a_B) = 3.0$ from the optimized quantum variational curve (Fig. 3A). We then obtain the 1st input value of $\varepsilon_r^c$ by applying Eq. 1, namely, $\varepsilon_r^c = a_B(\mu_o/m_o a_o)$ $= a_B/0.452$. The last expression was obtained by plugging $\mu_o = 0.117 m_o$ and $a_o = 0.0529 \, nm$ into Eq. 1. Thus, the 1st input value of $\varepsilon_r^c$ is 10.1. Consequently, $(\varepsilon_r^c/\varepsilon_r^s) = 10.1/2.5 \approx 4.0$ as we explained previously. The corresponding $C_1'$ value can be obtained by applying the Takagahara's graphical result of $A_1$ parameter (24) and Eq. 6, namely, $C_1' = (-A_1)/2 = +2.60/2 = 1.30$ as the 1st input value of $C_1'$. We then compute the 1st input values of the following two parameters to obtain the 1st output value of $(R_c/a_B)$: (i) $\xi_R' \equiv C_1' \gamma_c^2 = 1.30 \times (1.28)^2 = 2.130$ and (ii) $\nu \equiv 2(C_2 \gamma_c - \frac{1}{2})$ $= 2 \times (0.498 \times 1.28 - \frac{1}{2}) = 0.275$. We then obtain the following 1st output value of $(R_c/a_B)$ by plugging these two numerical values into Eq. 8: $(R_c/a_B) = 2.130 \times 0.407/0.275 = 3.15$. This 1st output value of $(R_c/a_B)$ is substantially different from the 1st input value of $(R_c/a_B) = 3.0$.

Accordingly, we adopt $(R_c/a_B) = 3.15$ as the 2nd input value in our self-consistent iterative calculations. We then obtain $\gamma_c = 1.27$ for $(R_c/a_B) = 3.15$ (Fig. 3A). Consequently, the 2nd input value of $a_B$ is 4.36 nm as $R_c = 13.75 \, nm$. We then obtain the 2nd input value of $\varepsilon_r^c$ by using $\varepsilon_r^c = a_B/0.452 = 9.65$. Thus, $(\varepsilon_r^c/\varepsilon_r^s) = 9.65/2.5 \approx 3.86$. The corresponding 2nd input value of $C_1'$ is given by $C_1' = (-A_1)/2 = +2.62/2 = 1.31$. We then calculate the 2nd input values of the following two parameters for $(R_c/a_B)$: (i) $\xi_R' \equiv C_1' \gamma_c^2 = 1.31 \times (1.27)^2 = 2.113$ and (ii) $\nu \equiv 2(C_2 \gamma_c - \frac{1}{2}) = 2 \times (0.498 \times 1.27 - \frac{1}{2}) = 0.265$. We obtain the following 2nd output value of $(R_c/a_B)$ by plugging these two numerical values into Eq. 8: $(R_c/a_B) = 2.113 \times 0.395/0.265 = 3.15$. Therefore, the 2nd output value is in good agreement with the 2nd input value of $(R_c/a_B)$. We then obtain the following converged values for colloidal MAPbBr$_3$ under the dielectric confinement: (i) $(R_c/a_B) = 3.15$, thus, $a_B = 4.36 \, nm$, (ii) $\varepsilon_r^c = 9.65$, (iii) $C_1' = 1.31$, (iv) $\gamma_c = 1.27$.

In the absence of dielectric confinement, we do not have to perform self-consistent iterative calculations. This is because $(R_c/a_B)$ has a fixed value of 2.45 [Eq. 10] in the absence of dielectric confinement, which would be valid for any photoluminescent single crystalline material, regardless of its chemical nature. As $R_c = 13.75 \, nm$ for MAPbBr$_3$ perovskite, $a_B$ is 5.61 nm for single crystalline MAPbBr$_3$ perovskite. We then obtain $\gamma_c = 1.36$ for $(R_c/a_B) = 2.45$ (Fig. 3A). Consequently, we obtain the following numerical values: (i) $\xi_R \equiv C_1 \gamma_c^2 = 1.786 \times (1.36)^2 = 3.304$ and (ii) $\nu \equiv$

$2\left(C_2\gamma_c - \frac{1}{2}\right) = 2 \times \left(0.498 \times 1.36 - \frac{1}{2}\right) = 0.355$. We compute the optimized $(R_c/a_B)$ by using these two values and Eq. 9: $(R_c/a_B) = 3.304 \times 0.263/0.355 = 2.45$. As expected, this precisely coincides with the prediction of Eq. 10, namely, $(R_c/a_B) = 2.45$. We obtain the optimized $\varepsilon_r^c$ value by using $\varepsilon_r^c = a_B/0.452 = 12.41$. Thus, we have the following optimized values for single crystalline MAPbBr$_3$ perovskite in the absence of the dielectric confinement: (i) $(R_c/a_B) = 2.45$, thus, $a_B = 5.61\ nm$, (ii) $\varepsilon_r^c = 12.41$, (iii) $C_1 = 1.786$, (iv) $\gamma_c = 1.36$.

30. J. Park, H. M. Jang, S. Kim, S. H. Cho, T.-W. Lee, Electroluminescence of perovskite nanocrystals with ligand engineering. *Trends in Chemistry* **2**, 837-849 (2020).

31. Y. Kayanuma, Wannier exciton in microcrystals. *Solid State Commun*. **59**, 405-408 (1986).

32. G. T. Einevoll, Confinement of excitons in quantum dots. *Phys. Rev. B* **45**, 3410-3417 (1992).

33. C. Kittel, in *Introduction to Solid State Physics* (John Wiley & Sons, Inc., 1996; 7$^{th}$ ed.), Chapter 3.

34. Y. Zhang, Y. Liu, X. Chen, Q. Wang, Controlled synthesis of $Ag_2S$ quantum dots and experimental determination of the exciton Bohr radius. *J. Phys. Chem. C* **118**, 4918-4923 (2014).

35. Q. Wang, X.-D. Liu, Y.-H. Qiu, K. Chen, L. Zhou, Q.-Q. Wang, Quantum confinement effect and exciton binding energy of layered perovskite nanoplatelets. *AIP Advances* **8**, 025108 (2018).

36. R. Romestain, G. Fishman, Excitonic wave function, correlation energy, exchange energy, and oscillator strength in a cubic quantum dot. *Phys. Rev. B* **49**, 1774-1781 (1994).

37. R. M. Martin in *Electronic Structure: Basic Theory and Practical Methods* (Cambridge University Press, 2004), Appendix F (pp. 499-511).

38. V. D'Innocenzo, G. Grancini, M. J. P. Alcocer, A. R. S. Kandada, S. D. Stranks, M. M. Lee, G. Lanzani, H. J. Snaith, A. Petrozza, Excitons versus free charges in organo-lead tri-halide perovskites. *Nat. Commun*. **5**, 3586 (2014).

39. Z. Yang, A. Surrente, K. Galkowski, N. Bruyant, D. K. Maude, A. A. Haghighirad, H. J. Snaith, P. Plochocka, R. J. Nicholas, Unraveling the exciton binding energy and the dielectric constant in single-crystal methylammonium lead triiodide perovskite. *J. Phys. Chem. Lett*. **8**, 1851-1855 (2017).

40. S. H. Oh, H. M. Jang, Two-dimensional thermodynamic theory of epitaxial $Pb(Zr,Ti)O_3$ thin films. *Phys. Rev. B* **62**, 14757-14765 (2000).

41. W. Zhang, D. Dai, X. Chen, X. Guo, J. Fan, Red shift in the photoluminescence of colloidal carbon quantum dots induced by photon reabsorption. *Appl. Phys. Lett.* **104**, 091902 (2014).

**Table 1.** Exciton Bohr radius ($a_B$), dielectric constant of NC core ($\varepsilon_r^c$), and exciton binding energy ($E_b$) of MAPbBr$_3$ and CsPbBr$_3$ with and without dielectric-confinement effect.

| Material parameters | Spherical MAPbBr$_3$ | | | | Cubic-shaped CsPbBr$_3$ | | | |
|---|---|---|---|---|---|---|---|---|
| | $a_B$ (nm) | $\varepsilon_r^c$ | $\gamma_c$ | $E_b$ (meV) | $a_B$ (nm) | $\varepsilon_r^c$ | $\gamma_c$ | $E_b$ (meV) |
| with dielectric confinement | 4.36 | 9.65 | 1.27 | 27.6 | 5.08 | 6.72 | 1.28 | 35.1 |
| without dielectric confinement | 5.61 | 12.41 | 1.36 | 22.6 | 6.00 | 7.94 | 1.33 | 30.7 |

**Table 2.** Effects of dielectric constant of dispersion medium ($\varepsilon_r^s$) on various material parameters that include exciton Bohr radius ($a_B$), dielectric constant of MAPbBr$_3$ NC core ($\varepsilon_r^c$), and effective Coulomb interaction coefficient ($C_1'$).

| $\varepsilon_r^s$ | $C_1'$ | $\varepsilon_r^c$ | $a_B$ (nm) | $\left(\dfrac{\varepsilon_r^c}{\varepsilon_r^s}\right)$ | $\gamma_c$ | $\alpha_c$ (nm) | $(\varepsilon_r^c \gamma_c a_B)^{-1}$ $\times 100$ (nm$^{-1}$) |
|---|---|---|---|---|---|---|---|
| 0.5 | 1.04 | 8.01 | 3.62 | 16.02 | 1.20 | 4.34 | 2.9 |
| 1.5 | 1.19 | 8.89 | 4.02 | 5.93 | 1.23 | 4.94 | 2.3 |
| 2.0 | 1.26 | 9.34 | 4.22 | 4.67 | 1.25 | 5.28 | 2.0 |
| 2.5 | 1.31 | 9.65 | 4.36 | 3.86 | 1.27 | 5.53 | 1.9 |
| 3.0 | 1.36 | 9.96 | 4.50 | 3.32 | 1.27 | 5.72 | 1.7 |
| 3.5 | 1.42 | 10.29 | 4.65 | 2.94 | 1.28 | 5.95 | 1.6 |
| 5.0 | 1.54 | 11.02 | 4.98 | 2.20 | 1.31 | 6.52 | 1.4 |
| 7.5 | 1.63 | 11.53 | 5.21 | 1.54 | 1.34 | 6.98 | 1.2 |
| 10.0 | 1.72 | 12.08 | 5.46 | 1.21 | 1.35 | 7.37 | 1.1 |
| 12.4 | 1.786 | 12.41 | 5.61 | 1.00 | 1.36 | 7.63 | 1.1 |

# * Figure Captions

**Fig. 1. Size-dependent PL-wavelength and linewidth of MAPbBr$_3$ NCs.** The PL wavelength (blue circles) is weakly dependent on the MAPbBr$_3$ PNC size down to the modal radius $R$ of ~5 nm but decreases rapidly below this critical value (left-hand side of the green vertical zone). Here, the dispersion medium is consisted of toluene ($\varepsilon_r = 2.4$) with a minor amount of oleic acid ($\varepsilon_r = 2.5$) as a capping agent (*13*), where $\varepsilon_r$ denotes the relative dielectric permittivity (dielectric constant). On the contrary, the corresponding PL linewidth (black squares; $\Gamma_{FWHM}$) increases rapidly with decreasing size below this critical radius of ~5 nm. ***Fig. 1** reproduced from (*12*) with permission. Data are practically identical to those reported in (*12*) but reexamined in this study with a different viewpoint.

**Fig. 2. Dispersed NCs and NC-size-dependent exciton energy for two distinct cases related to electron-hole (*e-h*) correlation.** **(A)** Colloidally dispersed NCs (with $\varepsilon_r^c$) in a medium having lower dielectric constant of $\varepsilon_r^s$, where the NC-medium interface is covered by the adsorbed capping-agent molecules such as oleic acid. On the other hand, an exciton in each spherical NC or QD (quantum dot) with the radius $R$ is schematically described in the right-hand-side diagram. The image charges $e'\ (= +R \cdot e/r_e)$ and $h'\ (= -R \cdot e/r_h = +R \cdot h/r_h)$ are located at the distances $r_e'\ (= R^2/r_e)$ and $r_h'\ (= R^2/r_h)$, respectively, from the center of the NC, where $r_e$ and $r_h$, respectively, denote the distances of the electron and the hole from the center. **(B)** The net exciton energy above the bulk band gap, $\Delta E_g$, in the absence of the internal *e-h* correlation effect (blue curve). It can be shown theoretically that $R_{eq} = 2R_c$ under this condition, where $R_{eq}$ is the equilibrium size of the spherical NC in which the exciton is thermodynamically stable satisfying the equilibrium condition that $\left(\partial \Delta E_g/\partial R\right)_{\varepsilon_r^c} = 0$. See the Supplementary Materials for details. Thus, $\Delta E_g = 0$ corresponds to the conduction band minimum (CBM). However, this diagram predicts a contradictory result because the PL red shift does not stop at $R_c$ and continues to occur until the NC size reaches $R_{eq}$. In other words, $\Delta E_g < 0$ for $R > R_c$. This is in direct contradiction to the observation that the PL red shift ceases at $R_c$. **(C)** The net exciton energy above the bulk band gap (blue curve) in the presence of the internal *e-h* correlation which is characterized by the parameter $\alpha$ in the exciton wave function (Eq. 4). It can be shown that $R_{eq}$ is equal to $R_c$ under this condition. **(D)** A hypothetical colloidal dispersion in which the NC cores having characteristic $R_c$ are dispersed in a dielectric medium with its dielectric constant $\varepsilon_r^s$ equal to that of the NC cores. Thus, it can be viewed as a photoluminescent single crystal. The dielectric-confinement effect disappears owing to the absence of dielectric boundary between the imaginary core (with $R_c$)

and the dielectric medium which is factually a single-crystalline matrix (green color).

**Fig. 3. Quantum variationally optimized curves for the *e-h* correlation parameter ($\alpha$) *versus* the NC size.** **(A)** Optimized $(\alpha/a_B)$ plotted as a function of $(R/a_B)$ for spherical NCs. Reproduced with permission from Kayanuma. *Solid State Commun.* **59**, 405 (1986). Copyright 1986 Pergamon Press (Elsevier Ltd). **(B)** Optimized $(\alpha/a_B)$ plotted as a function of $(L/2a_B)$, where $L$ is the edge length of a cubic-shaped NC. Reproduced with permission from Romestain and Fishman. *Phys. Rev. B* **49**, 1774-1781 (1994). Copyright 1994 American Physical Society.

**Fig. 4. PL emission spectra and related morphological data.** **(A)** Size-dependent PL emission spectra of MAPbBr$_3$ under an excitation of 405 nm (left-hand side) and TEM images with the corresponding size-distribution histograms at two selected modal diameters, $D_p = 9.9\ nm$ and $D_p = 23.7\ nm$ (right-hand side). A numerical value in each PL spectrum denotes the corresponding modal diameter ($D_p$). We observed that the peak wavelength in the PL emission spectrum ceases to show any further red shift at $D_p = 27.5\ nm$. Thus, the onset radius of the PL blue shift ($R_c$) is $13.75\ nm$ (= 27.5/2) as shown in Fig. 1. **(B)** Size-dependent PL emission spectra of CsPbBr$_3$ under an excitation of 405 nm (left-hand side) and TEM images at two selected edge lengths, $L = 10.6\ nm$ and $L = 14.7\ nm$ (right-hand side). A numerical value in each PL spectrum denotes the corresponding edge length ($L$) of cubic-shaped CsPbBr$_3$. We observed that the peak wavelength in the PL emission spectrum ceases to show any further red shift at $27.0\ nm$. Thus, the onset edge-length of the PL blue shift ($L_c$) is $27.0\ nm$. *****Fig. 4a** reproduced from (*12*) with permission. Data are practically identical to those reported in (*12*) but reexamined in this study using a new theoretical model on $(R_c/a_B)$.

**Fig. 5**. **Effect of the degree of dielectric confinement on important material parameters. (A)** Self-consistently estimated $\varepsilon_r^s$-dependent $\varepsilon_r^c$ and $C_1'$ values. Note that $C_1'$ reaches its maximum value of 1.786 (red circle) when $\varepsilon_r^c/\varepsilon_r^s = 1.0$ at $\varepsilon_r^s = 12.41$. **(B)** Self-consistently estimated $\varepsilon_r^s$-dependent $a_B$ and $(R_c/a_B)$ values. $(R_c/a_B)$ shows its universal value of 2.45 in the absence of dielectric confinement (single crystal) when $\varepsilon_r^s = \varepsilon_r^c = 12.41$. **(C)** A semiempirical linear correlation between $a_B$ and $(\varepsilon_r^c \gamma_c a_B)^{-1}$, suggesting a good linear relation between $a_B$ and $|E_{corr}|$, the Coulomb-like *e-h* correlation energy (Eq. 11).

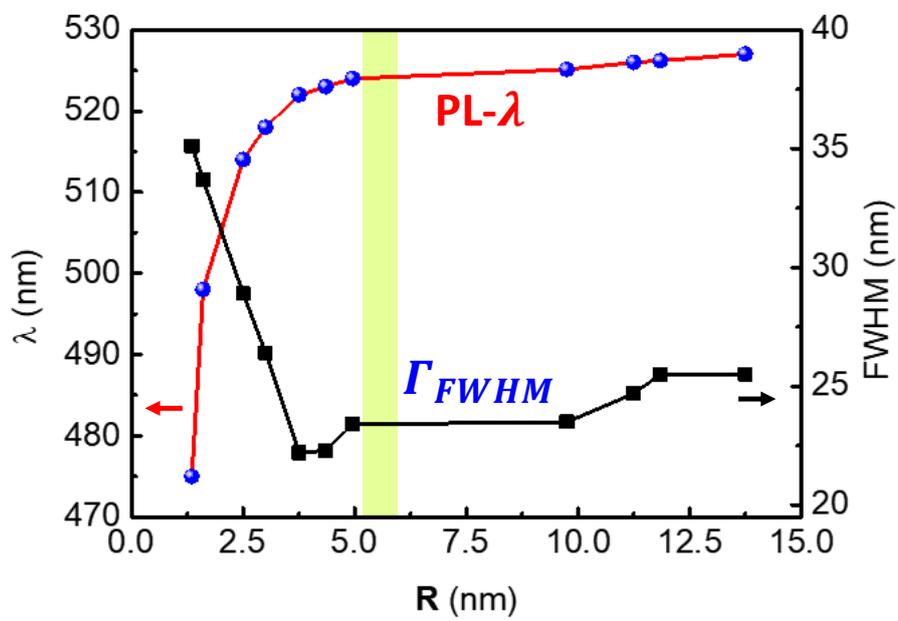

**Figure 1**

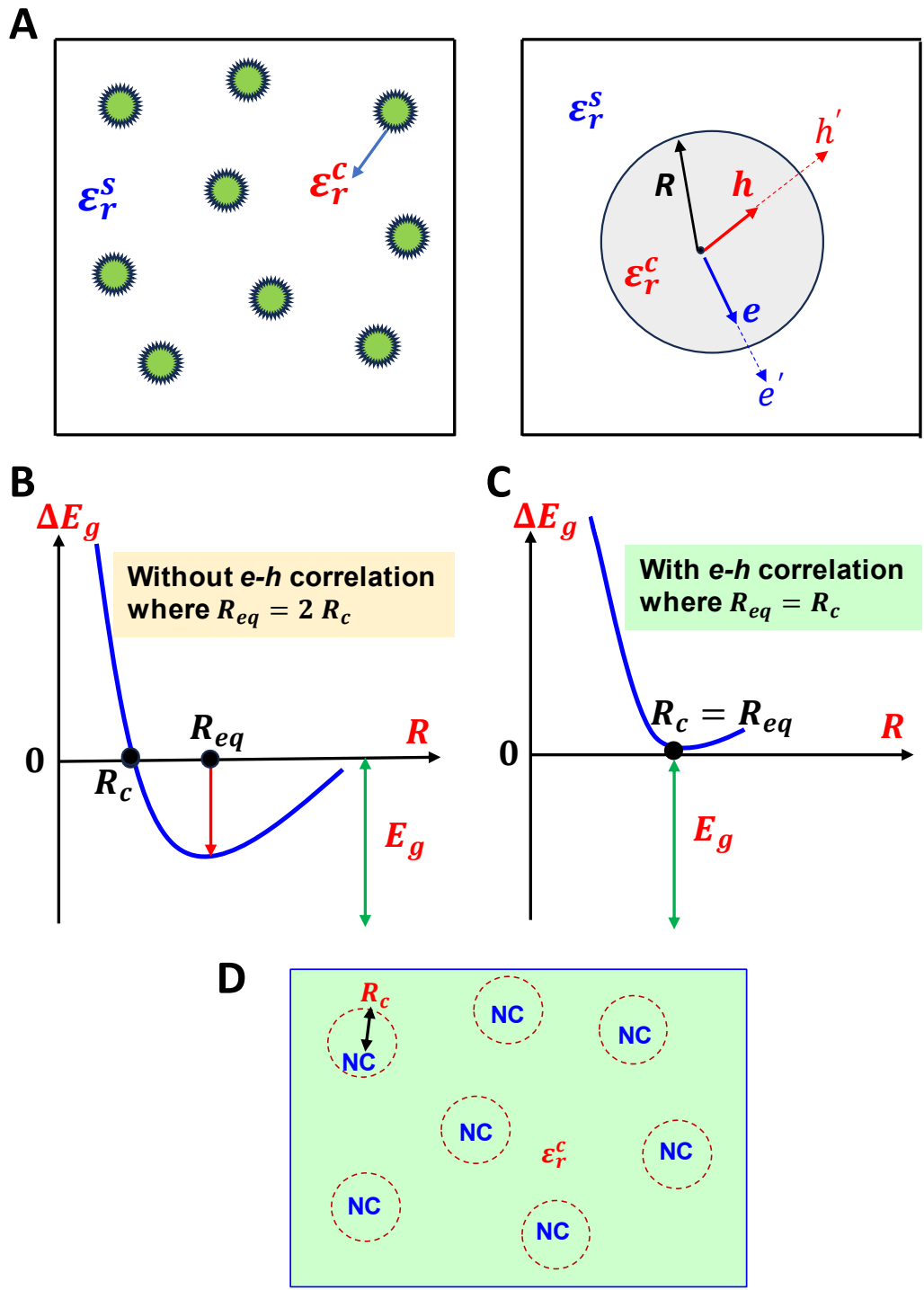

**Figure 2**

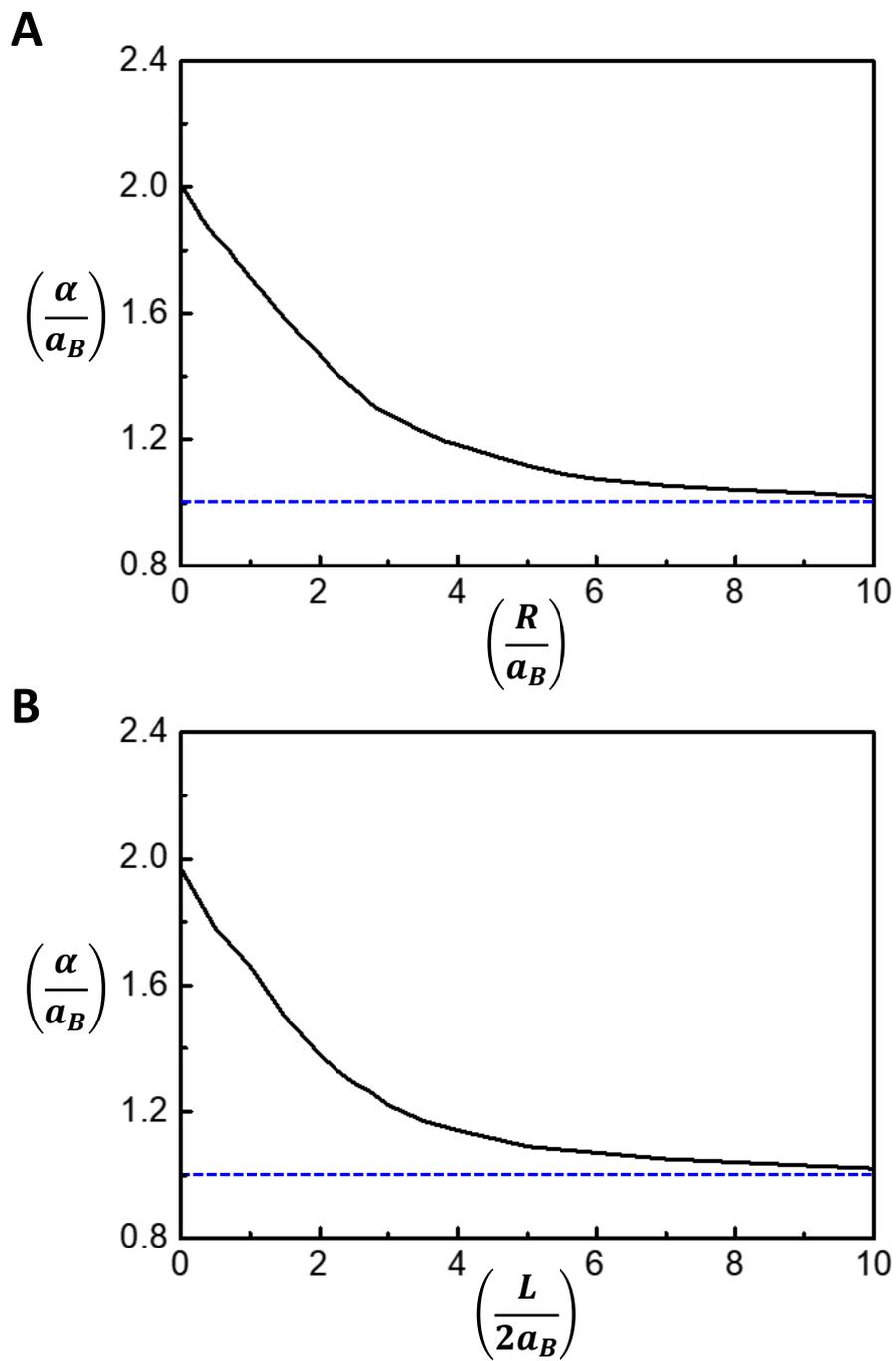

**Figure 3**

**A** (MAPbBr$_3$)

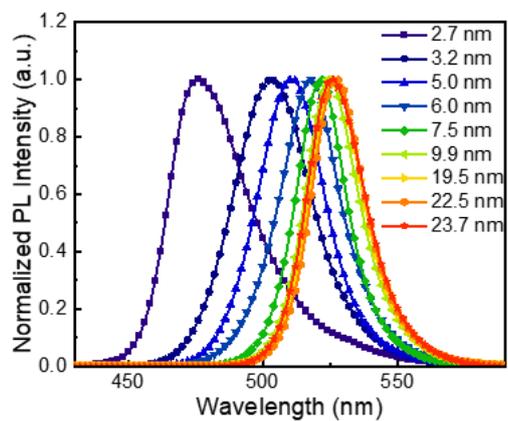
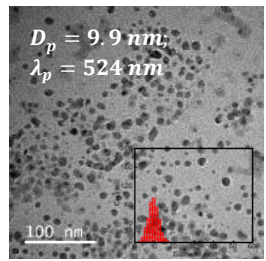
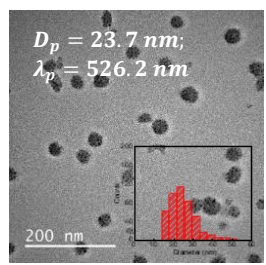

**B** (CsPbBr$_3$)

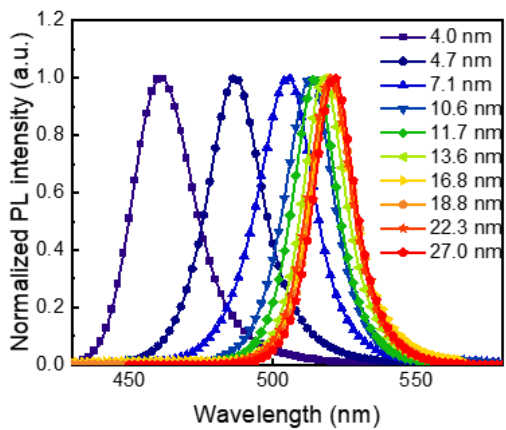
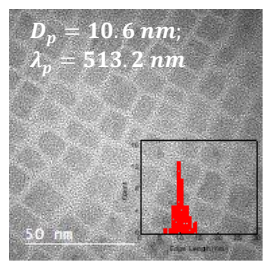
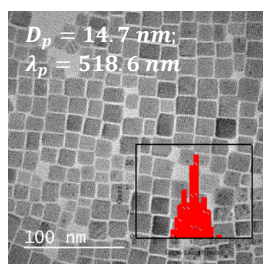

# Figure 4

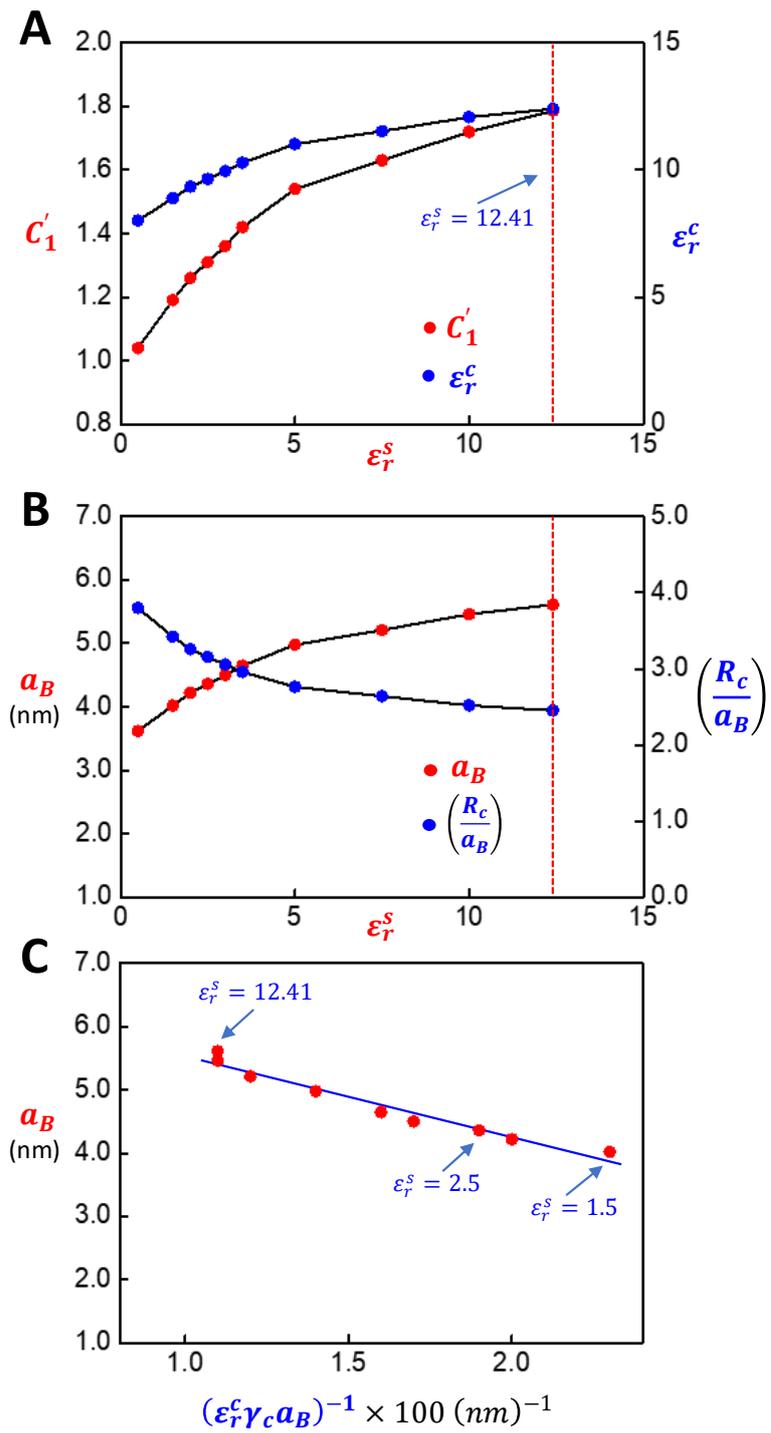

**Figure 5**

# Supplementary Materials for

## Exciton Bohr radius of lead halide perovskites for photovoltaic and light-emitting applications

Hyun Myung Jang*, Kyung Yeon Jang, Song Hee Lee, Jinwoo Park,
Tae-Woo Lee*

**\*** Corresponding author. Email: jang0227@snu.ac.kr (H.M.J.)

**The PDF file includes:**

(i) Thermodynamic stability at the onset size of PL blue shift

(ii) Kinetic energy of correlated excitons in a cubic-shaped nanocrystal

(iii) Coulombic interaction energy of correlated excitons in a cubic-shaped nanocrystal

### (i) Thermodynamic stability at the onset size of PL blue shift

The PL wavelength ($\lambda$) of a photoluminescent semiconductor increases until it approaches $\lambda$ of its critical size ($R_c$) for the onset of PL blue shift, where the corresponding band gap is equal to its bulk value, $E_{g(bulk)}$. The exciton energy at $R_c$ is equal to the conduction band minimum. In other words, the net exciton energy above the band gap [*i.e.*, $\Delta E_g(R;\alpha)$] is 0 at $R_c$. It is highly important to know whether this critical size corresponds to the thermodynamically stable size for the exciton that resides in a given NC. According to Eq. 3, $\Delta E_g(R;\alpha) = 0$ at $R_c$ predicts that $(R_c/a_B) = \pi^2/2C_1'$ in the absence of any internal *e-h* correlation effect (*i.e.*, $\alpha = 0$), where $a_B$ denotes the exciton Bohr radius (Eq. 1). On the other hand, $(\partial \Delta E_g/\partial R)_{\mu_o,\varepsilon_r^c} = 0$ using Eq. 3 leads to the following relation for the equilibrium NC size ($R_{eq}$) in which the exciton is thermodynamically stable: $(R_{eq}/a_B) = \pi^2/C_1' = 2(R_c/a_B)$. Thus, $R_{eq} = 2R_c$, as shown schematically in Fig. 2B. In such a case, however, one would expect a continuous PL red shift until the NC size is equal to $2R_c$ ($= R_{eq}$), which is in contradiction to experimental observations (S1, S2). Moreover, this violates the requirement that $\Delta E_g(R;\alpha) \geq 0$ with its minimum at $R_c$. This suggests that $R_c = R_{eq}$ for (i) $\Delta E_g(R;\alpha) \geq 0$ and (ii) the surcease of PL red shift with increasing NC size beyond $R_c$ (Fig. 2C).

Considering Eq. 3, we should include the internal *e-h* correlation effect (that is a non-zero $\alpha$) to obtain a correct NC size in which $R_c = R_{eq}$ with $\Delta E_g(R;\alpha) = 0$ (Fig. 2C). To this end, we first impose the equilibrium condition to Eq. 3 $[(\partial \Delta E_g/\partial R)_{\mu_o,\varepsilon_r^c} = 0]$. This yields:

$$\frac{\hbar^2}{2\mu_o}\left[\left\{\frac{-2\pi^2}{R^3} - \frac{2}{\alpha^3}\left(\frac{\partial \alpha}{\partial R}\right)\right\} + \left(\frac{2}{a_B}\right)\left\{\frac{C_1'}{R^2} + \frac{C_2}{\alpha^2}\left(\frac{\partial \alpha}{\partial R}\right)\right\}\right] = 0 \tag{S.1}$$

If we impose the condition that the bulk exciton Bohr radius $a_B$ remains unchanged under a finite change in $\alpha$ or $R$, we can obtain the following integral equation from Eq. S.1:

$$\int\left\{\left(\frac{2}{\alpha^3}\right) - \left(\frac{2}{a_B}\right)\frac{C_2}{\alpha^2}\right\}d\alpha = \int\left\{\left(\frac{-2\pi^2}{R^3}\right) + \left(\frac{2}{a_B}\right)\frac{C_1'}{R^2}\right\}dR \tag{S.2}$$

The solution of Eq. S.2 is given by

$$\left\{\frac{-1}{\alpha^2} + \left(\frac{2}{a_B}\right)\frac{C_2}{\alpha}\right\} = \left\{\frac{\pi^2}{R^2} - \left(\frac{2}{a_B}\right)\frac{C_1'}{R}\right\} + K \tag{S.3}$$

where $K$ is an integration constant. It can be shown that $K = 0$ by exploiting the fact that $C_2 = 0.498$ and $\alpha \to 1.0 a_B$ as $R \to \infty$ (Fig. 3A): $K + 0 = (-1/a_B^2) + (2/a_B)(C_2/a_B) \approx 0$. Eq. S.3 with $K = 0$ shows that $\alpha$ is not independent of $R$ and they are interrelated with each other. The relation between the two parameters at the critical size ($R_c$) can be obtained by rewriting the above equation with $K = 0$.

$$\frac{\alpha_c}{a_B} \equiv \gamma_c = \frac{C_2 \left(\frac{R_c}{a_B}\right)^2}{\left\{\pi^2 - 2C_1'\left(\frac{R_c}{a_B}\right)\right\}} \cdot \left\{1 - \sqrt{1 - \frac{\left(\pi^2 - 2C_1'\left(\frac{R_c}{a_B}\right)\right)}{(C_2)^2 \left(\frac{R_c}{a_B}\right)^2}}\right\} \quad (S.4)$$

According to Eq. S.4, there exists one unique value of the internal $e$-$h$ correlation parameter $\alpha_c$ at the critical size $R_c$ ($= R_{eq}$) for fixed values of $C_1', C_2, a_B$ (Fig. 3A). In the absence of dielectric confinement, $C_1' = C_1 = 1.786$. Under this condition, we can show that $(R_c/a_B) = 2.45$, which is universally valid for any photoluminescent semiconductor (Eq. 10 of the main manuscript). Plugging these values into Eq. S.4 predicts that $\gamma_c \equiv \alpha_c/a_B = 1.35$. This value is very close to 1.36 as obtained from the quantum variationally optimized curve at $(R_c/a_B) = 2.45$ in the absence of dielectric confinement (Fig. 3A and Table 1). Thus, the present prediction of $\alpha$ (i.e., Eq. S.4), which is based on thermodynamic equilibrium, is nearly consistent with the quantum variational prediction which is based on the numerical minimization of the energy expectation-integral (S3).

We obtained Eq. S.4 by applying the equilibrium condition of $\left(\partial \Delta E_g/\partial R\right)_{\mu_o, \varepsilon_r^c} = 0$ at $R_c$. However, we have not yet examined the necessary condition of the thermodynamic stability at $R_c$. To do this, we take the second partial derivative of $\Delta E_g$ with respect to $R$.

$$\left(\frac{\partial^2 \Delta E_g(R;\alpha)}{\partial R^2}\right)_\alpha = \frac{\hbar^2}{2\mu_o}\left[\left\{\frac{6\pi^2}{R^4} + \frac{6}{\alpha^4}\left(\frac{\partial \alpha}{\partial R}\right) - \frac{2}{\alpha^3}\left(\frac{\partial^2 \alpha}{\partial R^2}\right)\right\} - \left(\frac{2}{a_B}\right)\left\{\frac{2C_1'}{R^3} + \frac{2C_2}{\alpha^3}\left(\frac{\partial \alpha}{\partial R}\right) - \frac{C_2}{\alpha^2}\left(\frac{\partial^2 \alpha}{\partial R^2}\right)\right\}\right] \approx$$

$$\frac{\hbar^2}{2\mu_o}\left[\frac{6\pi^2}{R^4} - \frac{6S}{\alpha^4} - \left(\frac{2}{a_B}\right)\left\{\frac{2C_1'}{R^3} - \frac{S}{\alpha^3}\right\}\right] \approx +\frac{0.29\,\hbar^2}{2\mu_o a_B^4} > 0 \text{ at } R_c \quad (S.5)$$

We used a linear approximation in obtaining the third expression, $\left(\frac{\partial \alpha}{\partial R}\right) = -S$, where $S$ is a positive constant. We then extracted the following value from the quantum variationally optimized curve in Fig. 3A: $S \approx 0.10$ for $2.45 \leq (R/a_B) \leq 4.0$. We subsequently examined thermodynamic stability of exciton in a MAPbBr$_3$ NC under the dielectric confinement with $\varepsilon_r^S = 2.5$ (toluene-based dispersion medium). We computationally show that $R_c = 3.15 a_B$, $\alpha_c = 1.27 a_B$, $C_1' = 1.31$ for $\varepsilon_r^S = 2.5$. (See Materials and Methods). Using these values, we deduce that $\left(\partial^2 \Delta E_g/\partial R^2\right)_\alpha = +0.29\,\hbar^2/2\mu_o a_B^4 > 0$ at $R_c$ (Eq. S.5). This lucidly demonstrates the thermodynamic stability of exciton at $R_c$.

### (ii) Kinetic energy of correlated excitons in a cubic-shaped nanocrystal

The first step to theoretically predict the exciton Bohr radius and related properties in a cubic-shaped nanocrystal (NC) or quantum dot (QD) is the formulation of the kinetic energy of correlated excitons. For this, let us take the following product-type wave function ($\Psi$) for an exciton in a cubic QD:

$$\Psi = \Phi(r_e, r_h)\chi(r_{eh}) = \phi(r_e)\phi(r_h)\chi(r_{eh}), \quad (S.6)$$

where the degree of the *e-h* correlation is represented by a Slater-type correlation factor.

$$\chi(r_{eh}) = e^{-r_{eh}/\alpha} \tag{S.7}$$

This form of the *e-h* correlation gives rise to a closed form for the kinetic energy. As fully described in the main manuscript, the parameter $\alpha$ is a measure of the quantum *e-h* correlation with the asymptotic $\alpha$ of $\infty$ for a completely uncorrelated state (individually confined electron and hole).

Considering a sinusoidal form of the wave function for an electron (particle) in a cubic box, we can write the following form of the exciton wave function:

$$\Psi(r_e, r_h, r_{eh}) = N\, coskx_e coskx_h \cdot cosky_e cosky_h \cdot coskz_e coskz_h \cdot e^{-r_{eh}/\alpha}, \tag{S.8}$$

where the distance between electron and hole in an exciton ($r_{eh}$) is given by

$$r_{eh} = \sqrt{(x_e - x_h)^2 + (y_e - y_h)^2 + (z_e - z_h)^2} \tag{S.9}$$

Then, we consider the kinetic energy operator ($\tilde{T}$) defined by

$$\tilde{T} = \sum_{i=e,h} \sum_{q=x,y,z} -\frac{\hbar^2}{2m_i} \frac{\partial^2}{\partial q_i^2} = \sum_i -\frac{\hbar^2}{2m_i}\left(\frac{\partial^2}{\partial x_i^2} + \frac{\partial^2}{\partial y_i^2} + \frac{\partial^2}{\partial z_i^2}\right), \tag{S.10}$$

where the right-hand side expression is valid for a Cartesian coordinate. We further define the differential operator of the form:

$$\tilde{D} \equiv \frac{\partial}{\partial q_i} \tag{S.11}$$

We subsequently consider the following type of the expectation value integral ($W$) using the above differential operator:

$$W \equiv \langle \Psi | \tilde{D}^2 | \Psi \rangle = \left\langle \Psi \left| \frac{\partial^2}{\partial q_i^2} \right| \Psi \right\rangle \tag{S.12}$$

Eq. S.12 can be further expanded using Eq. S.6:

$$W = \langle \Phi\chi|\tilde{D}^2|\Phi\chi\rangle = \langle \Phi\chi|\tilde{D}\cdot\tilde{D}|\Phi\chi\rangle = \langle \Phi\chi|\tilde{D}|\chi\tilde{D}\Phi + \Phi\tilde{D}\chi\rangle =$$
$$\langle \Phi\chi|(\tilde{D}\Phi)(\tilde{D}\chi) + \chi\tilde{D}^2\Phi + (\tilde{D}\Phi)(\tilde{D}\chi) + \Phi\tilde{D}^2\chi\rangle = 2\langle \Phi\chi|(\tilde{D}\Phi)(\tilde{D}\chi)\rangle + \langle \Phi\chi^2|\tilde{D}^2\Phi\rangle +$$
$$\langle \Phi^2\chi|\tilde{D}^2\chi\rangle = 2A + \langle \Phi\chi^2|\tilde{D}^2\Phi\rangle + \langle \Phi^2\chi|\tilde{D}^2\chi\rangle, \tag{S.13}$$

where the integral $A$ is defined by

$$A \equiv \langle \Phi\chi|(\tilde{D}\Phi)(\tilde{D}\chi)\rangle = \langle(\Phi\tilde{D}\Phi)|(\chi\tilde{D}\chi)\rangle = \left\langle \left(\tfrac{1}{2}2\Phi\tilde{D}\Phi\right)\middle|\left(\tfrac{1}{2}2\chi\tilde{D}\chi\right)\right\rangle. \tag{S.14}$$

Thus, we obtain the following expression of $A$:

$$A = \left\langle \left(\tfrac{1}{2} 2\Phi \widetilde{D}\Phi\right) \middle| \left(\tfrac{1}{2} 2\chi \widetilde{D}\chi\right) \right\rangle = \tfrac{1}{4}\langle(\widetilde{D}\Phi^2)|(\widetilde{D}\chi^2)\rangle = \tfrac{1}{4}\langle(\widetilde{D}\Phi^2)(\widetilde{D}\chi^2)\rangle \tag{S.15}$$

Because $\cos(kx_e) = \cos\left(\frac{\pi}{L}\left(\pm\frac{L}{2}\right)\right) = \cos\left(\pm\frac{\pi}{2}\right) = 0$ at $q_i = \pm\frac{L}{2}$, $A$ can be further expanded to yield

$$A = \tfrac{1}{4}\langle(\widetilde{D}\Phi^2)(\widetilde{D}\chi^2)\rangle = \tfrac{1}{4}\iint_{q_i=-\infty}^{q_i=+\infty} \tfrac{\partial}{\partial q_i}\Phi^2(r_e, r_h)\tfrac{\partial}{\partial q_i}\chi^2(r_{eh})dq_i = \tfrac{1}{4}\iint(\widetilde{D}\chi^2)\left(\tfrac{\partial}{\partial q_i}\Phi^2\right)dq_i =$$
$$\tfrac{1}{4}\left[(\widetilde{D}\chi^2)(\Phi^2)\right]_{q_i=-\infty(=-L/2)}^{q_i=+\infty(=+L/2)} - \int \Phi^2(\widetilde{D}^2\chi^2)dq_i\right] = -\tfrac{1}{4}\int \Phi^2(\widetilde{D}^2\chi^2)dq_i \tag{S.16}$$

$\Phi = \Phi^*$ as $\Phi$ is real. Thus, $A$ can be further simplified to give

$$A = -\tfrac{1}{4}\int(\Phi^*)^2(\widetilde{D}^2\chi^2)dq_i = -\tfrac{1}{4}\langle\Phi^2|\widetilde{D}^2\chi^2\rangle \tag{S.17}$$

Combining Eq. S.17 with Eq. S.13 yield the following relation for the expectation value integral of the form $W \equiv \langle\Psi|\widetilde{D}^2|\Psi\rangle$ (Eq. S.12):

$$W = 2A + \langle\Phi\chi^2|\widetilde{D}^2\Phi\rangle + \langle\Phi^2\chi|\widetilde{D}^2\chi\rangle = \langle\Phi\chi^2|\widetilde{D}^2\Phi\rangle + \left\langle\Phi^2\middle|\chi(\widetilde{D}^2\chi) - \tfrac{1}{2}(\widetilde{D}^2\chi^2)\right\rangle \tag{S.18}$$

We now consider the expectation value of the exciton's kinetic energy $E_k$

$$\langle E_k\rangle = \langle\Psi|\widetilde{T}|\Psi\rangle = \langle\Phi\chi|\widetilde{T}|\Phi\chi\rangle, \tag{S.19}$$

where the kinetic energy operator $\widetilde{T}$ is

$$\widetilde{T} \equiv \sum_{i=e,h}\sum_{q=x,y,z} -\frac{\hbar^2}{2m_i}\frac{\partial^2}{\partial q_i^2} = \sum_{i=e,h}\sum_{q=x,y,z} -\frac{\hbar^2}{2m_i}\widetilde{D}^2 \tag{S.20}$$

Considering Eq. S.18 for the operator $\widetilde{D}^2$ and Eq. S.20 for the proportionality between $\widetilde{T}$ and $\widetilde{D}^2$, $\langle E_k\rangle$ in Eq. S.19 can be further expanded to yield

$$\langle E_k\rangle = \langle\Phi\chi|\widetilde{T}|\Phi\chi\rangle = \langle\Phi\chi^2|\widetilde{T}|\Phi\rangle + \left\langle\Phi^2\middle|\chi(\widetilde{T}\chi) - \tfrac{1}{2}(\widetilde{T}\chi^2)\right\rangle \tag{S.21}$$

We now consider the first term in the right-hand-side of Eq. S.21

$$\langle\Phi\chi^2|\widetilde{T}|\Phi\rangle = E_k\langle\Phi\chi^2|\Phi\rangle = E_k\langle\Phi\chi|\Phi\chi\rangle = E_k\langle\Psi|\Psi\rangle = E_k, \tag{S.22}$$

where $E_k$ denotes the eigenvalue of the kinetic energy. Thus, $E_k$ for a particle in a 3D box can be written as

$$E_k = \frac{\hbar^2}{2\mu}(k_x^2 + k_y^2 + k_z^2) = \frac{3\hbar^2 k^2}{2\mu} \tag{S.23}$$

The last expression is valid for a cubic QD, where $k_x = k_y = k_z = k$.

To begin with the second term in the right-hand-side of Eq. S.21, we separate the relative motion coordinate ($r = r_e - r_h$) from the center-of-mass coordinate where $R = M^{-1}(m_e r_e + m_h r_h)$. Then, the kinetic energy operator $\tilde{T}$ can be written as

$$\tilde{T} = \frac{P_R^2}{2M} + \frac{P_r^2}{2\mu} = \tilde{T}_R + \tilde{T}_r \tag{S.24}$$

Applying Eq. S.7 to Eq. S.24, we establish

$$\tilde{T}\chi = \tilde{T}_r \chi, \tag{S.25}$$

where the kinetic energy operator for the relative motion coordinate $\tilde{T}_r$ can be written as

$$\tilde{T}_r = \frac{P_r^2}{2\mu} = \frac{1}{2\mu}\left(-i\hbar \frac{\partial}{\partial r}\right)^2 = \frac{-\hbar^2}{2\mu}\frac{\partial^2}{\partial r^2} \tag{S.26}$$

Thus, $\tilde{T}_r \chi$ becomes

$$\tilde{T}_r \chi = \tilde{T}_r(e^{-r/\alpha}) = -\frac{\hbar^2}{2\mu}\frac{d}{dr}\left\{\left(-\frac{1}{\alpha}\right) e^{-\frac{r}{\alpha}}\right\} = -\frac{\hbar^2}{2\mu}\left(\frac{1}{\alpha}\right)^2 \chi \tag{S.27}$$

We consider the second term in the right-hand-side of Eq. S.21 using this result. This second term represents the *e-h* correlation contribution to the total kinetic energy of exciton. Then, applying Eq. S.25, the ket side of this second term in Eq. S.21 becomes

$$\chi(\tilde{T}\chi) - \frac{1}{2}(\tilde{T}\chi^2) = \chi \tilde{T}_r \chi - \frac{1}{2}\tilde{T}_r \chi^2 \tag{S.28}$$

The second term of Eq. S.28 proceeds further

$$-\frac{1}{2}\tilde{T}_r \chi^2 = \left(-\frac{1}{2}\right)\tilde{T}_r\{e^{-2r/\alpha}\} = \left(-\frac{1}{2}\right) \cdot -\frac{\hbar^2}{2\mu}\frac{d^2}{dr^2}\{e^{-2r/\alpha}\} = +\frac{\hbar^2}{2\mu}\left(\frac{2}{\alpha^2}\right)\chi^2 \tag{S.29}$$

On the other, the first term can be written by applying Eq. S.27 to $\chi \tilde{T}_r \chi$

$$\chi \tilde{T}_r \chi = -\frac{\hbar^2}{2\mu}\left(\frac{1}{\alpha}\right)^2 \chi^2 \tag{S.30}$$

Substituting Eqs. S.29 and S.30 into Eq. S.28 yields

$$\chi(\tilde{T}\chi) - \frac{1}{2}(\tilde{T}\chi^2) = -\frac{\hbar^2}{2\mu}\left(\frac{1}{\alpha}\right)^2 \chi^2 + \frac{\hbar^2}{2\mu}\left(\frac{2}{\alpha^2}\right)\chi^2 = +\frac{\hbar^2}{2\mu}\left(\frac{1}{\alpha^2}\right)\chi^2 \tag{S.31}$$

Thus, the second term of Eq. S.21 becomes

$$\left\langle \Phi^2 \middle| \chi(\tilde{T}\chi) - \frac{1}{2}(\tilde{T}\chi^2) \right\rangle = +\frac{\hbar^2}{2\mu}\left(\frac{1}{\alpha^2}\right)\langle \Phi^2 | \chi^2 \rangle = +\frac{\hbar^2}{2\mu}\left(\frac{1}{\alpha^2}\right)\langle \Phi\chi | \Phi\chi \rangle = +\frac{\hbar^2}{2\mu}\left(\frac{1}{\alpha}\right)^2 \quad \text{(S.32)}$$

Then, substituting Eqs. S.22, S.23 and S.32 into the right-hand side of Eq. S.21 yields

$$\langle E_k \rangle = \frac{\hbar^2}{2\mu}(k_x^2 + k_y^2 + k_z^2) + \frac{\hbar^2}{2\mu}\left(\frac{1}{\alpha}\right)^2 = +\frac{3\hbar^2 k^2}{2\mu} + \frac{\hbar^2}{2\mu}\left(\frac{1}{\alpha}\right)^2, \quad \text{(S.33)}$$

where $k_x = k_y = k_z = k = \frac{\pi}{L}$. Thus, $\langle E_k \rangle$ can be rewritten in terms of the edge length $L$ as

$$\langle E_k \rangle = \frac{3\hbar^2 \pi^2}{2\mu L^2} + \frac{\hbar^2}{2\mu}\left(\frac{1}{\alpha}\right)^2 \quad \text{(S.34)}$$

This relation is given in Eq. 12 of the main manuscript.

### (iii) Coulombic interaction energy of correlated excitons in a cubic-shaped nanocrystal

In this section, we theoretically derive the expectation value integral of the Coulomb-like attractive interaction $\langle V_c \rangle$ in terms of the ratio of two distinct three-fold integrals on the three relative coordinates between electron and hole, namely, $k(x_e - x_h)$, $k(y_e - y_h)$, and $k(z_e - z_h)$, where $k = \pi/L$. For this, we first write the Hamiltonian operator as the sum of two contributions:

$$\tilde{\mathcal{H}} = \tilde{\mathcal{H}}_k + \tilde{\mathcal{H}}_c(r_e, r_h) = \tilde{T} + \tilde{\mathcal{H}}_c(r_e, r_h), \quad \text{(S.35)}$$

where $\tilde{\mathcal{H}}_c$ is the Coulomb operator given by

$$\tilde{\mathcal{H}}_c = -\frac{e^2}{4\pi\varepsilon_o \varepsilon_r^c |r_e - r_h|} = -\frac{e^2}{4\pi\varepsilon_o \varepsilon_r^c \, r_{eh}} \quad \text{(S.36)}$$

We then define the expectation value of the Coulomb interaction.

$$\langle V_c \rangle = \int \Psi^*(r_e, r_h)\, \tilde{\mathcal{H}}_c \, \Psi(r_e, r_h)\, d^3 r_e d^3 r_h \quad \text{(S.37)}$$

Substituting Eq. S.7 for $\chi(r_{eh})$, Eq. S.8 for $\Psi(r_e, r_h)$, and Eq. S.9 for $r_{eh}$ into Eq. S.37 yields

$$\langle V_c \rangle = N^2 \iiint \iiint_{-L/2}^{+L/2} \cos^2 kx_e \cos^2 kx_h \cdot \cos^2 ky_e \cos^2 ky_h \cdot \cos^2 kz_e \cos^2 kz_h \cdot$$
$$\exp\left\{-\frac{2}{\alpha}\sqrt{(x_e - x_h)^2 + (y_e - y_h)^2 + (z_e - z_h)^2}\right\} \tilde{\mathcal{H}}_c \, dx_e dx_h dy_e dy_h dz_e dz_h \quad \text{(S.38)}$$

We define $kx_e \equiv \xi_e$ and $kx_h \equiv \xi_h$, where $k = \pi/L$. Then, consider the variation in the integral limit associated with this definition.

$$-\frac{L}{2} \leq x_e \leq +\frac{L}{2} \quad \Rightarrow \quad -\frac{\pi}{2} \leq \xi_e \leq +\frac{\pi}{2} \tag{S.39}$$

We also define the following two relations because the eigenfunction $\Psi(r_e, r_h)$ contains a function $[f(|\boldsymbol{r_e} - \boldsymbol{r_h}|)]$ with the inter-distance, $|\boldsymbol{r_e} - \boldsymbol{r_h}|$, as its independent variable:

$$\xi \equiv \xi_e - \xi_h \quad and \quad \xi' \equiv \xi_e + \xi_h \tag{S.40}$$

Then, define the following 1:1 correspondence: $\xi_1 \Leftrightarrow x, \; \xi_2 \Leftrightarrow y, \; and \; \xi_3 \Leftrightarrow z$.

We should calculate three integrals of the form for $x, y,$ and $z$ to evaluate $\langle V_c \rangle$:

$$I = \iint_{-\frac{L}{2}}^{+\frac{L}{2}} dx_e dx_h \cos^2(kx_e)\cos^2(kx_h) \cdot f(|\boldsymbol{r_e} - \boldsymbol{r_h}|) = \frac{1}{k^2}\int_{-\frac{\pi}{2}}^{+\frac{\pi}{2}} d\xi_e \int_{-\frac{\pi}{2}}^{+\frac{\pi}{2}} d\xi_h \cos^2(\xi_e)\cos^2(\xi_h) \cdot$$

$$f(|\xi_e - \xi_h|) = \frac{1}{4k^2}\int_0^\pi d\xi \left[(\pi - \xi)\{2 + \cos(2\xi)\} + \frac{3}{2}\sin(2\xi)\right] \cdot f(\xi) = \frac{1}{4k^2}\int_0^\pi d\xi \, g(\xi) f(\xi) \tag{S.41}$$

where $g(\xi)$ is defined as

$$g(\xi) \equiv (\pi - \xi)\{2 + \cos(2\xi)\} + \frac{3}{2}\sin(2\xi) \tag{S.42}$$

Thus, the six-fold integral of $\langle V_c \rangle$ [Eq. S.38] can be reduced to a three-fold integral by introducing the relative variable, $\xi(=\xi_x) \equiv \xi_e - \xi_h = k(x_e - x_h)$. In carrying out the integral of Eq. (S.41), we transformed $x_i$ to $\xi_i$, where $i = e$ or $h$. This transformation has a $\frac{1}{2}$ Jacobian, namely, $dx_e dx_h \Rightarrow \frac{1}{2} d\xi d\xi'$. In view of the last form of Eq. (S.41), the six-fold integral [Eq. (S.38)] can be written as

$$I_n(\beta) = 2^{-6} \int_0^\pi \int_0^\pi \int_0^\pi d\xi_1 \, d\xi_2 d\xi_3 g(\xi_1)g(\xi_2)g(\xi_3) \cdot \exp\left(-\frac{2\rho}{\beta}\right)\rho^n \tag{S.43}$$

where $\beta \equiv k\alpha$ and $\rho = kr(\xi_1, \xi_2, \xi_3)$. Since $\xi_x = k(x_e - x_h)$, we have the following relation for a three-dimensional relative variable:

$$\rho_{eh} = |\rho_e - \rho_h| = k|r_e - r_h| \tag{S.44}$$

Thus, $r/\alpha$ is given by $r/\alpha = kr/k\alpha = \rho/\beta$. Using these relations, Eq. (S.43) becomes

$$I_n(\beta) = 2^{-6} \int_0^\pi \int_0^\pi \int_0^\pi d\xi_x \, d\xi_y d\xi_z g(\xi_x)g(\xi_y)g(\xi_z) \cdot \exp(-2r/\alpha)(kr)^n \tag{S.45}$$

We further have: $\xi_x = kx, \; \xi_y = ky$ and $\xi_z = kz$. Thus, we obtain:

$$kr = k\sqrt{x^2 + y^2 + z^2} = \sqrt{\xi_x^2 + \xi_y^2 + \xi_z^2} \tag{S.46}$$

Substituting this relation into Eq. (S.45) yields

$$I_n(\beta) = 2^{-6} \int_0^\pi \int_0^\pi \int_0^\pi d\xi_x \, d\xi_y \, d\xi_z \, g(\xi_x) g(\xi_y) g(\xi_z) \cdot \exp\left\{-\frac{2}{k\alpha}\sqrt{\xi_x^2 + \xi_y^2 + \xi_z^2}\right\} (\xi_x^2 + \xi_y^2 + \xi_z^2)^{n/2} \quad (S.47)$$

where $\xi_x = \xi_{ex} - \xi_{hx} = kx_e - kx_h = k(x_e - x_h)$, $\xi_y = k(y_e - y_h)$ and $\xi_z = k(z_e - z_h)$. Using Eqs. (S.47), (S.36) and (S.38), it can be shown that the expectation value of the Coulomb interaction energy in a cubic-shaped nanocrystal or quantum dot is given by

$$\langle V_c \rangle = -\frac{e^2 k}{4\pi\varepsilon_o \varepsilon_r^c} \frac{I_{-1}(\beta)}{I_0(\beta)} = -\frac{e^2 \beta}{4\pi\varepsilon_o \varepsilon_r^c \alpha} \frac{I_{-1}(\beta)}{I_0(\beta)} = -\frac{e^2}{4\pi\varepsilon_o \varepsilon_r^c}\left(\frac{\pi}{L}\right)\frac{I_{-1}(\beta)}{I_0(\beta)} \quad (S.48)$$

In obtaining the second expression of Eq. (S.48), we used the relation that $k = \beta/\alpha$ [Eq. (S.43)]. On the other hand, we used the relation that $k = \pi/L$ [Eq. (S.33)] in obtaining the last expression of Eq. (S.48).

## References cited:

bibliography**S1.** Y. Zhang, Y. Liu, X. Chen, Q. Wang, Controlled synthesis of Ag$_2$S quantum dots and experimental determination of the exciton Bohr radius. *J. Phys. Chem. C* **118**, 4918-4923 (2014).

**S2.** Y.-H. Kim, C. Wolf, Y.-T. Kim, H. Cho, W. Kwon, S. Do, A. Sadhanala, C. G. Park, S.-W. Rhee, S. H. Im, R, Friend, T.-W. Lee, Highly efficient light-emitting diodes of colloidal metal-halide perovskite nanocrystals beyond quantum size. *ACS Nano* **11**, 6586-6593 (2017).

**S3.** Y. Kayanuma, Wannier exciton in microcrystals. *Solid State Commun*. **59**, 405-408 (1986).